# Sodium Incorporation in $CsPbBr_{3-x}I_x$ Nanocrystal Electrodes: Lattice Contraction and the Suppression of Field-Driven Iodine Expulsion

*Arun Kumar*[†], *Monojit Bag*[†, ‡, *].

†Advanced Research in Electrochemical Impedance Spectroscopy Laboratory, Indian Institute of Technology Roorkee, Roorkee 247667, India

‡Centre for Nanotechnology, Indian Institute of Technology Roorkee, Roorkee 247667, India

## Abstract

Mixed-halide perovskite nanocrystal electrodes fail in supercapacitors through field-driven halide segregation. In undoped $CsPbBr_2I$ this appears as a capacitance that climbs to 218% of its first-cycle value by cycle 1378 and then collapses to 39% by cycle 2500, with complete loss of the iodine signal from the cycled electrode. This work tests whether sodium incorporation suppresses that failure mode. Na-doped $CsPbBr_{3-x}I_x$ (x = 0, 1, 2) nanocrystals were prepared by ligand-assisted reprecipitation at a Na/Pb precursor ratio of 1.25:1.00 and compared with undoped analogues in 0.1 M tetrabutylammonium tetrafluoroborate in anhydrous dichloromethane. Sodium contracts the pseudocubic lattice parameter of $CsPbBr_3$ from 5.908 ± 0.029 to 5.851 ± 0.019 Å after correction for specimen displacement. Specific capacitance at 0.3 A $g^{-1}$ rises for every composition, from 42 to 75, from 63 to 96 and from 56 to 84.5 F $g^{-1}$. Na-$CsPbBr_2I$ gives the lowest charge-transfer resistance at 175 Ω and the highest ion diffusion coefficient at $1.6 \times 10^{-16}$ $m^2$ $s^{-1}$, and power-law exponents between 0.33 and 0.45 at all potentials examined show that charge storage is limited by ion transport through the pore network rather than by the interfacial process. In the sodium-containing (Na-$CsPbBr_2I$) electrode the capacitance rise reaches only 115% at cycle 600, no collapse follows, and 97% is retained at 2500 cycles, the excess above the first-cycle value being reduced by a factor of 7.9. Iodine is retained at unchanged binding energy, and the Br:I ratio measured by elemental mapping is 2.30 after cycling against 2.25 before.

## 1 Introduction

All-inorganic lead halide perovskite nanocrystals combine mixed ionic and electronic conductivity, a bandgap tunable across the visible range by halide substitution, and a defect population that is tolerated electronically rather than acting as deep recombination centres.[1–3] Colloidal $CsPbX_3$ (X= Cl, Br, I) nanocrystals with edge lengths of 4–15 nm were first prepared by hot injection, giving emission tunable between 410 and 700 nm at quantum yields of 50–90%.[1] Room-temperature routes followed, of which ligand-assisted reprecipitation is the most widely used because it requires neither inert atmosphere nor elevated temperature.[4,5] A broader survey of synthesis routes and the resulting optical and structural properties is given elsewhere.[6] The features that make these materials attractive optically — a soft, ionic lattice with low vacancy formation energies — are the same features that make ion migration facile. The same lability permits rapid post-synthetic anion exchange between compositions at room temperature.[7,8] Under sustained electrochemical bias it is ion migration, not electronic behaviour, that governs the response of the material. Halide perovskite supercapacitor electrodes have been reported over the past five years, and the capacitance values span two orders of magnitude depending on composition and electrolyte. Rod-shaped $CsPbBr_3$ in aqueous electrolyte delivers 121 F $g^{-1}$ at 5 mV $s^{-1}$ with 73% retention after 5000 cycles.[9] $CsPbBr_3$ nanocrystals in tetrabutylammonium hexafluorophosphate dissolved in dichloromethane store charge by a double-layer mechanism, reaching 528 mF $g^{-1}$ but retaining 90% of capacitance after 10 000 cycles at 100 mA $g^{-1}$.[10] Thin-film $CsPbBr_3$ symmetric devices give 17 mF $cm^{-2}$ with 89% retention over 5000 cycles, and the layered analogue $CsPb_2Br_5$ reaches 886 F $g^{-1}$ in three-electrode configuration with 76% retention over the same interval.[11] Lead-free $Cs_3Bi_2Cl_9$ films give 64 mF $cm^{-2}$ at retention near 90%.[12] Halide perovskites have also been examined as insertion electrodes for lithium-ion cells, and the broader case for their use in electrochemical energy storage has been reviewed recently.[13,14] The spread arises because the electrolyte determines whether charge is stored capacitively at the surface or through halide redox chemistry in the bulk, and because retention is quoted over cycle counts differing by a factor of four between studies.[15] Comparison of absolute capacitance across this literature is therefore of limited diagnostic value, whereas comparison of failure behaviour under a fixed protocol is informative. Halide segregation is the dominant failure mode of mixed-halide compositions. Illumination of $MAPb(I_{1-x}Br_x)_3$ produces a red-shifted photoluminescence peak at 1.68 eV with increased sub-bandgap absorption, and diffraction shows splitting into an iodide-rich and a bromide-rich phase.[16] The process reverses in the dark, which established it as compositional redistribution rather than decomposition. Segregation also proceeds under an applied electric field without carrier injection, so the driving force is the field acting on mobile ions rather than photogenerated carriers.[17] Reversible structural transitions of this kind have been observed directly by diffraction in perovskite nanocrystals rather than in thin films alone.[18] That distinction matters for an electrode operated at bias in the dark, where the field is present continuously and reverses polarity each cycle. Vacancy-mediated migration accounts for the ion selectivity: in $MAPbI_3$ the calculated barrier for iodide vacancy migration is 0.58 eV, matching a kinetic measurement of approximately 0.60 eV, while methylammonium migration requires 0.84 eV and lead is effectively immobile.[19,20] Bromide migrates more slowly than iodide because the smaller anion is more tightly bound within the octahedral cage, so in a mixed Br/I lattice the iodide sublattice moves first. Nanocrystals aggravate the problem, since halide vacancies at $CsPbBr_3$ surfaces migrate with lower barriers than bulk vacancies and the average barrier falls as grain size decreases.[21] The consequences in an operating supercapacitor electrode were quantified in earlier work from this laboratory.[22] Porous electrodes of $CsPbBr_{3-x}I_x$ (x = 0, 1, 2) prepared by ligand-assisted reprecipitation were cycled in tetrabutylammonium tetrafluoroborate dissolved in dichloromethane. $CsPbBr_3$ degraded monotonically, and $CsPbBrI_2$ collapsed within the first few hundred cycles while losing both halides. $CsPbBr_2I$ behaved unlike either: its capacitance climbed continuously to above 200% of the first-cycle value by cycle 1350, then fell to approximately 40% by cycle 2500. X-ray photoelectron spectroscopy of the cycled electrode showed the I 3d signal absent entirely; diffraction showed CsI and $PbBr_2$ alongside residual perovskite after 1250 cycles, with only $Cs_4PbBr_6$ and $PbBr_2$ remaining at 2500 cycles. Nitrogen adsorption linked the capacitance rise to an increase in pore volume as iodine left the lattice. The rise is therefore a symptom of the material consuming itself, not a measure of improving performance. An electrode that gains capacitance over a thousand cycles because its lattice is opening up will lose that capacitance once the lattice can no longer support itself. Read as a diagnostic signature rather than as an anomaly requiring explanation, the rise-and-collapse curve becomes a test: any treatment that suppresses field-driven halide migration should remove the rise, and its removal is then measurable evidence that migration has been suppressed. Monovalent alkali cations have been used to stabilise halide perovskites against segregation in optical and photovoltaic contexts. Potassium incorporation in mixed-halide films immobilises excess halide at grain boundaries, mitigates photoinduced segregation, and raises external luminescence yield.[23,24] Partial substitution of the A-site by caesium in mixed-cation formulations likewise improves thermal stability and reduces phase impurity.[25] Sodium incorporation in $CsPbBr_3$ nanocrystals raises photoluminescence quantum yield, and emission of the doped nanocrystals is retained at approximately 80% of its initial value after seven heating and cooling cycles against a larger decay for undoped material; the optimum occurred at a Na/Pb precursor ratio of 1.25, with energy-dispersive spectroscopy giving an incorporated sodium content near 0.5 at%.[26] Addition of $NaBH_4$ during synthesis contracts the lattice measurably, the (100) spacing of $FAPb(Br/Cl)_3$ falling from 6.09 to 5.78 Å, a contraction of 0.31 Å accompanied by a shift of the reflection from 14.53° to 15.31°.[27] In each of these studies stability was assessed optically, thermally, or under illumination. None tested the doped material under sustained electrochemical bias, where the field is applied deliberately, reverses polarity, and acts on an electrode already in contact with a liquid electrolyte capable of dissolving expelled halide. Where the sodium ion sits in the lattice remains unsettled, and the two candidate sites predict opposite effects on halide mobility. Density functional calculations on doped $CsPbBr_3$ give a formation energy of 0.35 eV for substitution at the lead site against 1.73 eV at the caesium site, favouring B-site occupancy;[28] the sodium doping study of $CsPbBr_3$ nanocrystals reaches the same conclusion and reports that sodium at the lead site raises the diffusion barrier of the bromide vacancy.[26] The lattice-contraction argument drawn from diffraction data is framed instead as A-site substitution of the larger caesium ion.[27] Effective ionic radii make the ambiguity concrete: $Na^+$ is 1.02 Å in sixfold coordination, close to 1.19 Å for $Pb^{2+}$ in the same coordination, but far below 1.88 Å for $Cs^+$ in the twelvefold coordination of the perovskite A-site.[29] A-site substitution is isovalent and requires no charge compensation. B-site substitution replaces a divalent cation with a monovalent one and must be compensated, most plausibly by halide vacancies, which would raise the vacancy concentration and act against the intended suppression of halide transport. B-site substitution also carries a larger formation energy than A- or X-site substitution across the $CsPbX_3$ family, so incorporation is expected to be partial and dilute.[30] The site question determines whether contraction and vacancy compensation reinforce or oppose one another. Solid-state nuclear magnetic resonance on alkali-doped mixed-cation perovskites shows that small cations such as $K^+$ and $Rb^+$ may not enter the three-dimensional lattice at all, segregating instead into secondary phases.[31] Two gaps follow from this body of work. Alkali incorporation has not been tested under the conditions of an operating supercapacitor electrode, as distinct from the optical and photovoltaic conditions in which the strategy was developed. And no measurement has connected a doping-induced structural change to the electrochemical failure signature it is intended to remove. This work addresses both. Sodium was incorporated into $CsPbBr_3$, $CsPbBr_2I$ and $CsPbBrI_2$ nanocrystals during ligand-assisted reprecipitation at a Na/Pb precursor ratio of 1.25:1.00, and the doped and undoped series were compared under an identical electrochemical protocol. Diffraction establishes the phase and tracks the lattice parameter across the halide series, and transmission electron microscopy establishes particle size and habit. X-ray photoelectron spectroscopy and energy-dispersive mapping recorded before and after cycling establish the chemical state of lead and the retention of iodine in the cycled electrode, the measurement on which the undoped material failed most conspicuously. Cyclic voltammetry, galvanostatic charge–discharge and impedance spectroscopy establish the capacitive response and the interfacial kinetics, and the comparison of capacitance retention over 2500 cycles tests the diagnostic signature directly. Sodium incorporation contracts the lattice, lowers the charge-transfer resistance, and removes the rise-and-collapse behaviour of the undoped iodide-containing composition, while iodine is retained in the cycled electrode at unchanged binding energy.

## 2. Experimental Section

### 2.1 Materials

Caesium bromide (CsBr, 99.9%), caesium iodide (CsI, 99.9%), lead(II) bromide ($PbBr_2$, 99.99%), lead(II) iodide ($PbI_2$, 99.99%) and sodium bromide (NaBr, 99.99%) were obtained from TCI Chemicals and Sigma-Aldrich. Oleic acid (OA, 90%), oleylamine (OAm, 70%), anhydrous *N*,*N*-dimethylformamide (DMF, 99.8%), anhydrous toluene (99.8%) and *N*-methyl-2-pyrrolidone (NMP, 99%) were obtained from SRL Chemicals. Poly(vinylidene fluoride) (PVDF) and Super P conductive carbon were obtained from Sigma-Aldrich, and tetrabutylammonium tetrafluoroborate (TBTF, 98%) and anhydrous dichloromethane (DCM, 99.5) from SRL Chemicals. All chemicals were used as received.

### 2.2 Synthesis of undoped and $Na^+$-doped $CsPbBr_{3-x}I_x$ nanocrystals

Nanocrystals were prepared at room temperature by ligand-assisted reprecipitation, following the route of Li et al. with modifications.[5]

*Base precursor (solution A).* For $CsPbBr_3$, CsBr and $PbBr_2$ were combined in a 1:1 molar ratio, 0.4 mmol of each, and dissolved in 10 mL of anhydrous DMF. Oleylamine and oleic acid were added in a 1:2 volume ratio. The mixture was stirred at 500 rpm at 50 °C for 15 min and then sonicated for 15 min, and the two steps were repeated until a clear solution was obtained. For the mixed-halide compositions the caesium and lead contents were held constant and the $PbBr_2$:$PbI_2$ molar ratio was varied to give $CsPbBr_2I$ and $CsPbBrI_2$, with CsI substituted for CsBr in the corresponding proportion.

*Dopant precursor (solution B).* NaBr, 1 mmol, was dissolved in 1 mL of anhydrous DMF under constant stirring.

*Doping and crystallisation.* Solution B, 50 µL, was combined with 1 mL of solution A to give a Na/Pb atomic ratio of 1.25:1.00. This ratio was selected because it gave the largest gain in emission stability in Na-doped $CsPbBr_3$ nanocrystals.[26] The combined precursor, 100 µL, was added dropwise to 1 mL of anhydrous toluene under vigorous stirring, on which the nanocrystals precipitated. Large aggregates and unreacted precursor were discarded, and the supernatant was centrifuged at 10,000 rpm for 10 min. The recovered powder was dried under vacuum at 70°C for 12 h. Undoped controls were prepared in the same session by the same procedure with solution B omitted and all other quantities unchanged.

**2.3 Electrode fabrication**

Active material, Super P conductive carbon and PVDF were combined in a mass ratio of 70:15:15 and dispersed in NMP to give a homogeneous slurry, which was stirred for 24 h. The slurry was coated onto graphite sheet of $1 \times 1$ $cm^2$ active area and dried at 80 °C under vacuum for 24 h.

**2.4 Characterisation**

X-ray diffraction patterns were recorded on a Bruker D8 Advance diffractometer with Cu Kα radiation ($\lambda = 1.5406$ Å) over 10–50° 2θ. Transmission electron micrographs were recorded on a JEOL JEM-3200FS microscope, with samples drop-cast from toluene dispersion onto carbon-coated copper grids. Particle-size distributions were obtained by measuring multiple particles per sample from images taken in several distinct fields. Surface morphology and elemental mapping were performed on a Carl Zeiss Gemini field-emission scanning electron microscope with an attached energy-dispersive X-ray spectrometer. X-ray photoelectron spectra were recorded on a PHI VersaProbe III spectrometer with monochromatic Al Kα radiation (1486.6 eV), and binding energies were referenced to the adventitious C 1s peak at 284.8 eV.[32]

**2.5 Electrochemical measurements**

All measurements were made in a three-electrode cell containing 0.1 M tetrabutylammonium tetrafluoroborate (TBTF) in anhydrous dichloromethane, with the coated graphite as working electrode, Ag/AgCl as reference and platinum as counter electrode, on an Autolab potentiostat. Cyclic voltammograms were recorded at scan rates from 5 to 140 mV $s^{-1}$. Galvanostatic charge–discharge curves were recorded at current densities from 0.2 to 0.7 A $g^{-1}$. Impedance spectra were recorded from 0.01 Hz to 100 kHz, at open-circuit potential and at DC bias values of 0, 0.2, 0.4, 0.6, 0.8 and 1.0 V against Ag/AgCl. Cycling stability was measured over 2500 galvanostatic charge–discharge cycles. Specific capacitance, areal capacitance, energy density and power density were calculated using Equations S1–S6, and ionic conductivity and diffusion coefficients from the impedance data using the Bandara–Mellander treatment given in Equations S7–S12.[33]

## 3 Result and Discussion

### 3.1 Rational Design of Sodium-Stabilised Perovskite Nanocrystals

Undoped mixed-halide perovskite electrodes fail through a sequence that begins with halide transport rather than with electronic degradation. Under an applied field the iodide sublattice moves first, because the barrier for vacancy-mediated iodide migration is lower than that for the other constituent ions; in $MAPbI_3$ the calculated value is 0.58 eV for iodide against 0.84 eV for the A-site cation, with lead effectively immobile.[19] Local separation into iodide-rich and bromide-rich regions follows, and in an electrode immersed in a liquid electrolyte the expelled iodine is removed from the particle surface rather than remaining available for the reverse process. In earlier work on undoped $CsPbBr_{3-x}I_x$ electrodes the sequence was tracked to completion: the I 3d photoemission signal disappeared entirely, CsI and $PbBr_2$ appeared in the diffraction pattern, and the pore volume increased as the lattice opened.[22] The design strategy adopted here targets the first step in that sequence. Contracting the unit cell reduces the free volume of the migration pathway between adjacent halide sites, which raises the barrier that a migrating halide ion must overcome. Incorporation of a smaller cation than either of the native cations provides such a contraction. $Na^+$ has an effective ionic radius of 1.02 Å in sixfold coordination, against 1.19 Å for $Pb^{2+}$ in the same coordination and 1.88 Å for $Cs^+$ in the twelvefold coordination of the perovskite A-site.[29] Substitution at either site therefore contracts the lattice, and the contraction has been observed directly in nanocrystals grown with a sodium source, where the (100) spacing of $FAPb(Br/Cl)_3$ fell from 6.09 to 5.78 Å.[27] Partial replacement of $Pb^{2+}$ by smaller divalent cations produces the same effect, with the lattice parameter of $CsPb_{1-x}M_xBr_3$ contracting linearly with dopant fraction.[34,35] Which site the sodium ion occupies is not settled, and the two possibilities carry different consequences. First-principles calculations on $CsPbBr_3$ give a formation energy of 0.35 eV for substitution at the lead site against 1.73 eV at the caesium site,[28] and the same preference is reported alongside an increase in the bromide vacancy diffusion barrier for sodium at the lead site.[26] Substitution at the caesium site is isovalent and needs no compensating defect. Substitution at the lead site replaces a divalent with a monovalent cation and requires charge compensation, most plausibly by halide vacancies, which would raise the vacancy population even while the lattice contracts. Formation energies for B-site substitution across the $CsPbX_3$ family are larger than for the A- or X-sites, so incorporation is expected to be dilute.[30,36] The two effects — a narrower migration pathway and a possible increase in vacancy concentration — act in opposite directions, and the experimental measurements reported below constrain their net outcome without resolving the site assignment itself. Figure 1 sets out the synthetic route and the electrode architecture. Nanocrystals were grown by ligand-assisted reprecipitation at room temperature,[4,5] with sodium introduced through the precursor solution at a Na/Pb ratio of 1.25:1.00, the ratio at which the largest gain in thermal stability was reported for $CsPbBr_3$ nanocrystals.[26] That ratio was held constant across all three compositions, so the results below describe the halide dependence at a single doping level.

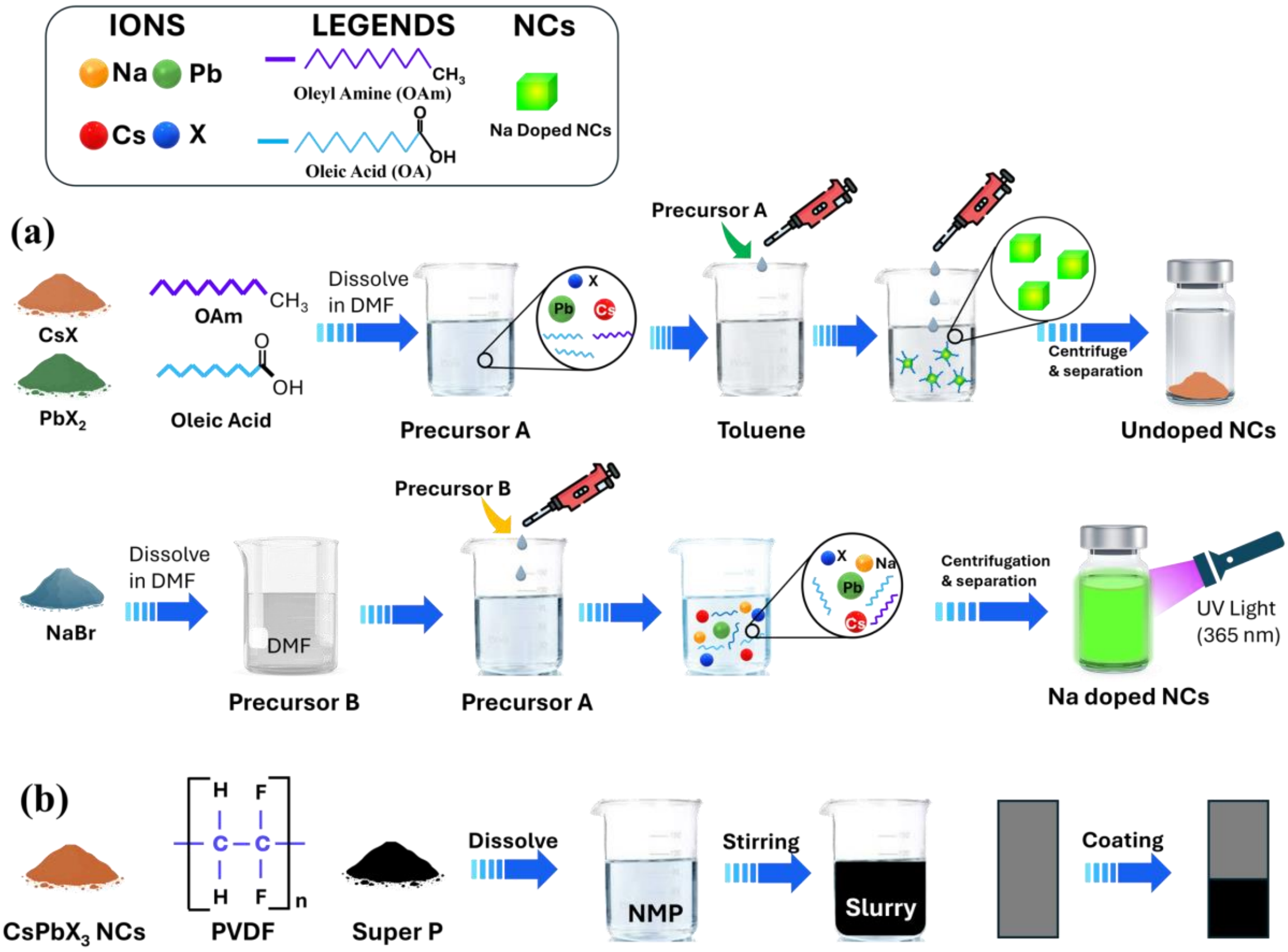

**Figure 1.** Synthesis route and electrode architecture. (a) Ligand-assisted reprecipitation of Na-doped $CsPbBr_{3-x}I_x$ (x = 0, 1, 2) nanocrystals. Precursor solution A contains CsX and $PbX_2$ with oleic acid and oleylamine in dimethylformamide; solution B contains NaBr in dimethylformamide, added at a Na/Pb ratio of 1.25:1.00. 100 µL of the combined precursor was injected into 1 mL anhydrous toluene under vigorous stirring, and the nanocrystals were recovered by centrifugation at 10 000 rpm and dried under vacuum. (b) Electrode fabrication: nanocrystals, Super P conductive carbon and poly(vinylidene fluoride) in the mass ratio 70:15:15, dispersed in N-methyl-2-pyrrolidone and cast onto graphite of 1 × 1 $cm^2$ active area, then dried at 80 °C under vacuum for 24 h.

### 3.2 Structure, Lattice Contraction and Morphology

**Phase and lattice parameter.** Figure 2(a) shows the region of the diffraction pattern containing the strongest reflection of each composition, with the undoped and sodium-containing materials plotted in pairs. All six patterns index to the orthorhombic perovskite structure in space group *Pbnm*, the room-temperature phase of $CsPbBr_3$.[37] The strongest reflection in the undoped $CsPbBr_3$ pattern, at 30.35°, is indexed as (220) and corresponds to the (200) reflection of the pseudocubic parent cell; the (004) reflection calculated at 30.44° lies within the observed line width of 0.27° and is not resolved from it. A weaker reflection at 28.79° is indexed as (212). This reflection carries half-integer pseudocubic indices and arises from octahedral tilting, so its presence establishes that the nanocrystals adopt the orthorhombic rather than the cubic structure at room temperature. The strongest reflection moves to higher angle in every sodium-containing sample. For $CsPbBr_3$ the (220) reflection moves from 30.35 to 31.42°, for $CsPbBr_2I$ from 30.18 to 30.37°, and for $CsPbBrI_2$ from 29.63 to 29.80°. The corresponding pseudocubic lattice parameters, refined against the peak positions with a specimen-displacement correction, are 5.908 and 5.851 Å for $CsPbBr_3$ and its sodium-containing analogue, 5.918 and 5.882 Å for $CsPbBr_2I$, and 6.025 and 5.992 Å for $CsPbBrI_2$, corresponding to contractions of 0.96, 0.61 and 0.56% respectively. The orthorhombic distortion is not resolved at the observed line width, so the pseudocubic parameter rather than the individual cell edges is reported. The direction of the shift is consistent with replacement of a larger by a smaller cation.[27,29,34] Across the undoped series the pseudocubic parameter increases monotonically with iodide content, from 5.908 Å for $CsPbBr_3$ to 5.918 Å for $CsPbBr_2I$ and 6.025 Å for $CsPbBrI_2$, following the larger ionic radius of iodide. The sodium-containing series follows the same trend displaced to smaller values, at 5.851, 5.882 and 5.992 Å, which indicates that sodium incorporation and halide substitution act on the lattice independently rather than competing. The pattern of Na-$CsPbBr_2I$ in Figure S1, contains reflections at 20.20, 22.55, 28.79, 39.09, 42.63 and 45.86° that do not index to the perovskite. These are attributed to a secondary phase formed under the bromide-rich conditions created by addition of the sodium bromide precursor, the zero-dimensional phase $Cs_4PbBr_6$ being the most likely candidate on the basis of its reported reflection positions.[38,39] Phase identification from peak position alone is not conclusive for a reference pattern of this line density, and quantitative phase analysis is not attempted here. The perovskite reflections of this composition remain the dominant feature of the pattern. The electrochemical response reported below for this composition is therefore that of a mixture in which the perovskite is the dominant crystalline phase rather than the only one.

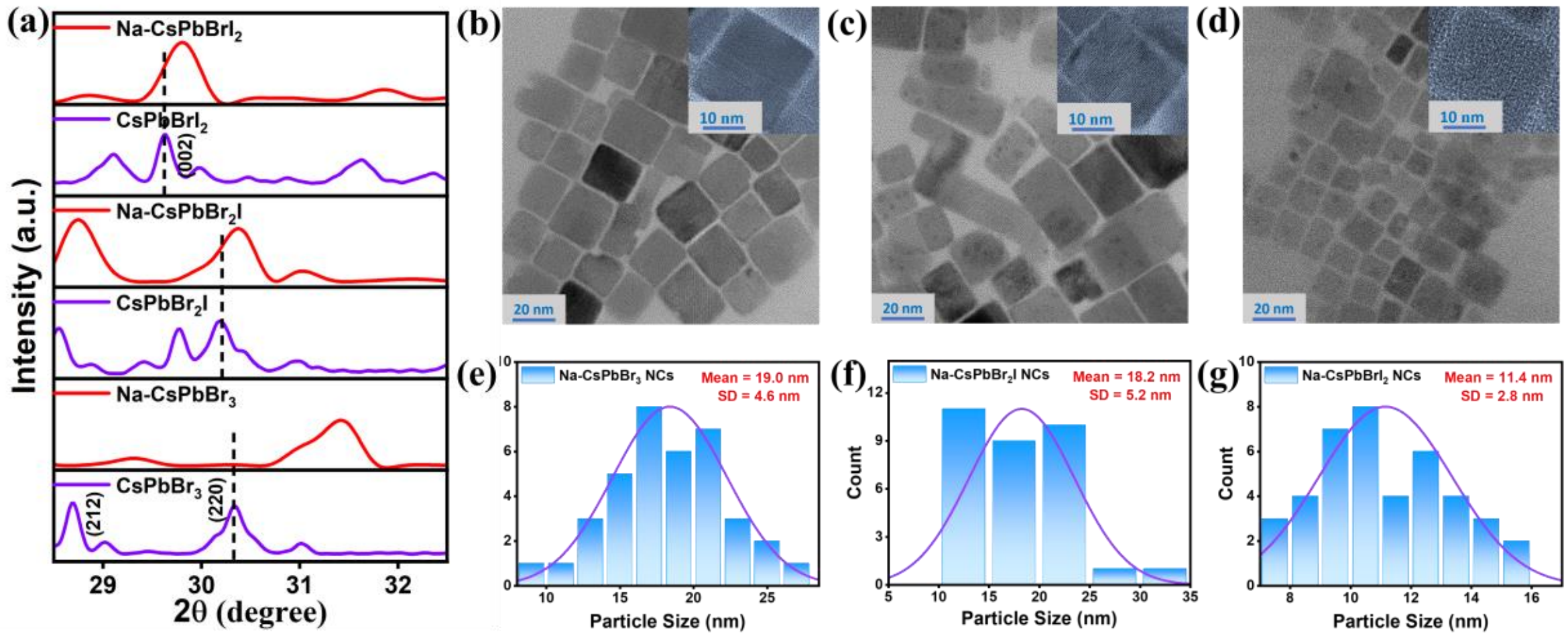

**Figure 2.** Structure and morphology of Na-doped $CsPbBr_{3-x}I_x$ nanocrystals. (a) X-ray diffraction patterns over 28.5–32.5° for undoped and Na-doped $CsPbBr_3$, $CsPbBr_2I$ and $CsPbBrI_2$, recorded with Cu Kα radiation (λ = 0.15406 nm) on a Bruker D8 Advance diffractometer; dashed lines mark the position of the strongest reflection in each undoped pattern. Traces are offset vertically for clarity and intensities are not on a common scale. (b–d) Transmission electron micrographs of Na-$CsPbBr_3$, Na-$CsPbBr_2I$ and Na-$CsPbBrI_2$ recorded on a JEM-3200FS microscope; scale bars 20 nm. Insets show high-resolution images of single particles; scale bars 10 nm. (e–g) Particle-size distributions for the same three samples, with Gaussian fits. Mean edge length and standard deviation are given on each panel.

**Morphology and size.** Figure 2(b–d) shows transmission electron micrographs of the three sodium-containing materials. All three consist of discrete, well-separated cuboidal particles with square projections, the habit characteristic of nanocrystals grown by reprecipitation in the presence of oleic acid and oleylamine.[1,5] The high-resolution insets show continuous lattice fringes extending across each particle, so the particles are single-crystalline and sodium incorporation at this level does not introduce visible structural disorder. Particle-size distributions measured from sample are given in Figure 2(e–g). Na-$CsPbBr_3$ has a mean edge length of 19.0 ± 4.6 nm and Na-$CsPbBr_2I$ of 18.2 ± 5.2 nm, while Na-$CsPbBrI_2$ is smaller at 11.4 ± 2.8 nm. Micrographs and size distributions of the undoped materials are given in our previous study for comparison.[22] The reduction in particle size on moving to the iodide-rich composition matters for the electrochemical results that follow, since it increases the specific surface area available for double-layer formation while also increasing the fraction of halide ions residing at or near a surface, where migration barriers are lower than in the bulk.[21] The two effects act on capacitance and on stability in opposite directions, and Section 3.5 returns to this point when the retention of the three compositions is compared.

### 3.3 Surface and Chemical State of the Cycled Electrode

**Surface and elemental changes on cycling.** Figure 3 collects the surface and chemical-state characterisation of the Na-$CsPbBr_2I$ electrode before and after 2500 galvanostatic cycles at 0.5 A $g^{-1}$; the equivalent data for Na-$CsPbBr_3$ and Na-$CsPbBrI_2$ are given in Figures S2-S5, and survey spectra for all three compositions in Figure S6. The pristine electrode consists of aggregated nanocrystal clusters distributed across the graphite substrate, with individual features below 200 nm resolved in the inset of Figure 3(a). After cycling the surface has coarsened into larger, more rounded particles and the fine texture of the pristine electrode is no longer visible (Figure 3b), a change that follows from repeated solvation and redeposition at the particle surface during cycling in a non-aqueous electrolyte[40,41] and that reduces the accessible area without requiring loss of the perovskite phase. Whether the composition changed alongside the morphology is answered by the elemental maps recorded from the same regions. Bromine and iodine are distributed uniformly across the pristine surface with no measurable enrichment or depletion at the micrometre scale, and both remain uniform after cycling; more usefully, the ratio of bromine to iodine coverage is 2.25 before cycling and 2.30 after, unchanged within measurement scatter. Preferential loss of one halide would raise that ratio, since iodide is the mobile species under bias,[19] so the coarsening seen in the micrographs is a rearrangement of the surface rather than a stripping of iodine from it. Absolute map density is not compared between the two measurements because it depends on acquisition time and display scaling.

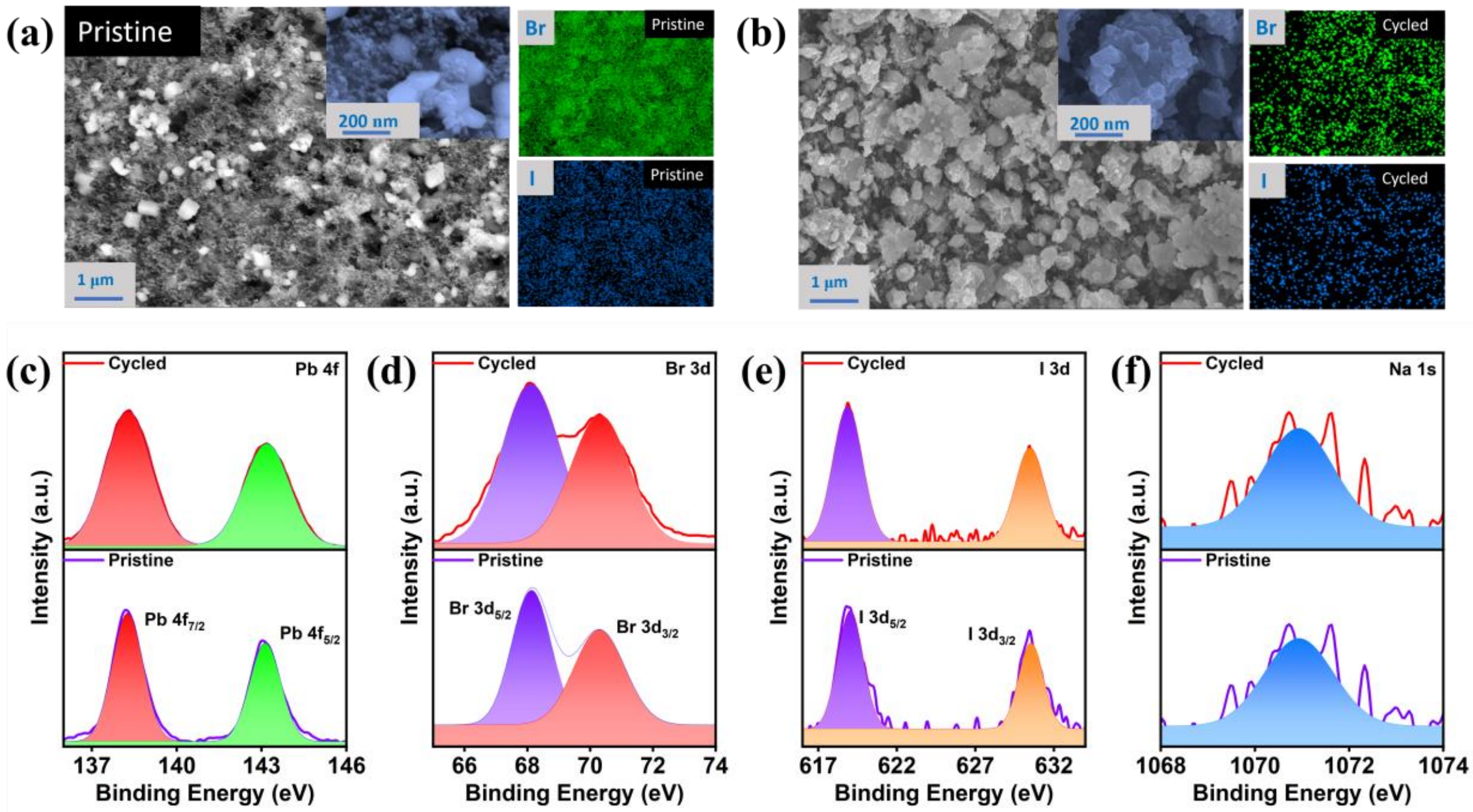

**Figure 3.** Surface morphology, elemental distribution and chemical state of the Na-$CsPbBr_2I$ electrode before and after cycling. (a) Field-emission scanning electron micrograph of the pristine electrode with higher-magnification inset, and energy-dispersive X-ray maps of Br and I from the same region. (b) The same measurements on an electrode after 2500 galvanostatic charge–discharge cycles in 0.1 M tetrabutylammonium tetrafluoroborate in anhydrous dichloromethane. Scale bars 1 μm; insets 200 nm. (c–f) High-resolution X-ray photoelectron spectra of (c) Pb 4f, (d) Br 3d, (e) I 3d and (f) Na 1s for the pristine (lower trace) and cycled (upper trace) electrode, recorded referenced to the C 1s peak at 284.8 eV. Shaded areas are fitted components.

**Chemical state.** The Pb 4f spectrum in Figure 3(c) shows the $4f_{7/2}$ and $4f_{5/2}$ components at 138.5 and 143.4 eV, separated by the spin–orbit splitting of 4.9 eV and at binding energies characteristic of $Pb^{2+}$ in the halide perovskite lattice.[42,43] No component appears at lower binding energy after cycling, so lead is not reduced to the metallic state under the conditions applied here, and since metallic lead is the endpoint of several proposed degradation routes in halide perovskites its absence sets a limit on how far reduction proceeds at the electrode surface. Bromine likewise retains both spin–orbit components after cycling, at 68.0 and 70.2 eV (Figure 3d). Both the Br 3d and Pb 4f envelopes are broader in the cycled electrode than in the pristine one, indicating a wider distribution of chemical environments and matching the partial surface reconstruction seen in the micrographs. The I 3d spectrum in Figure 3(e) separates the doped material from the undoped case most sharply. Both the $3d_{5/2}$ and $3d_{3/2}$ components are present after cycling at 619.0 and 630.5 eV, unchanged from the pristine electrode and with a flat baseline between them, so iodine is retained at the surface in the chemical state in which it began. In undoped $CsPbBr_2I$ cycled under the same protocol the I 3d signal was absent from the corresponding spectrum.[22] Sodium incorporation therefore prevents the loss of surface iodine that accompanied the capacitance collapse in the undoped material, and the Na 1s signal in Figure 3(f) shows that the dopant itself is retained rather than leached into the electrolyte during cycling. Its intensity reflects the dilute incorporation expected from the formation energies for substitution at either cation site,[28,30] and the spectrum does not by itself distinguish which site is occupied. The incorporated sodium concentration follows from the nominal precursor ratio and the energy-dispersive estimate in Table S1; inductively coupled plasma spectroscopy was not performed, so the concentration is not established independently of these two. Sodium retention is what the proposed mechanism requires, since a dopant that left the lattice within the first hundred cycles could not suppress halide migration over 2500. These measurements establish that the cycled surface retains lead in the divalent state, both halides in their original proportion, and sodium.

### 3.4 Capacitive Response of the Sodium-Containing Electrodes

**Charge storage behaviour.** Figure 4 presents the electrochemical response of the three Na-doped electrodes in 0.1 M tetrabutylammonium tetrafluoroborate (TBTF) in anhydrous dichloromethane, measured in a three-electrode cell against Ag/AgCl. The galvanostatic profiles at 0.2 A $g^{-1}$ in Figure 4(a) are near-triangular with a small departure from linearity, which indicates that charge is stored both in the electrical double layer and through a diffusion-limited faradaic contribution at the particle surface.[44,45] Discharge times differ markedly across the series, and so do the accessible potential windows: −0.60 to 0.80 V for Na-$CsPbBr_3$, −0.60 to 0.80 V for Na-$CsPbBr_2I$ and −0.65 to 0.60 V for Na-$CsPbBrI_2$. The narrower window of the iodide-rich composition limits its energy density independently of its capacitance. Voltammograms at 5 mV $s^{-1}$ (Figure 4d) are quasi-rectangular and free of resolved redox peaks, the response expected when double-layer storage dominates,[46,47] and Na-$CsPbBr_2I$ encloses the largest area of the three. The GCD profiles of all three doped materials recorded at current densities of 0.2–0.7 A $g^{-1}$ and the CV profiles obtained at scan rates of 5–140 mV $s^{-1}$ are provided in Figure S7 and S8, respectively.

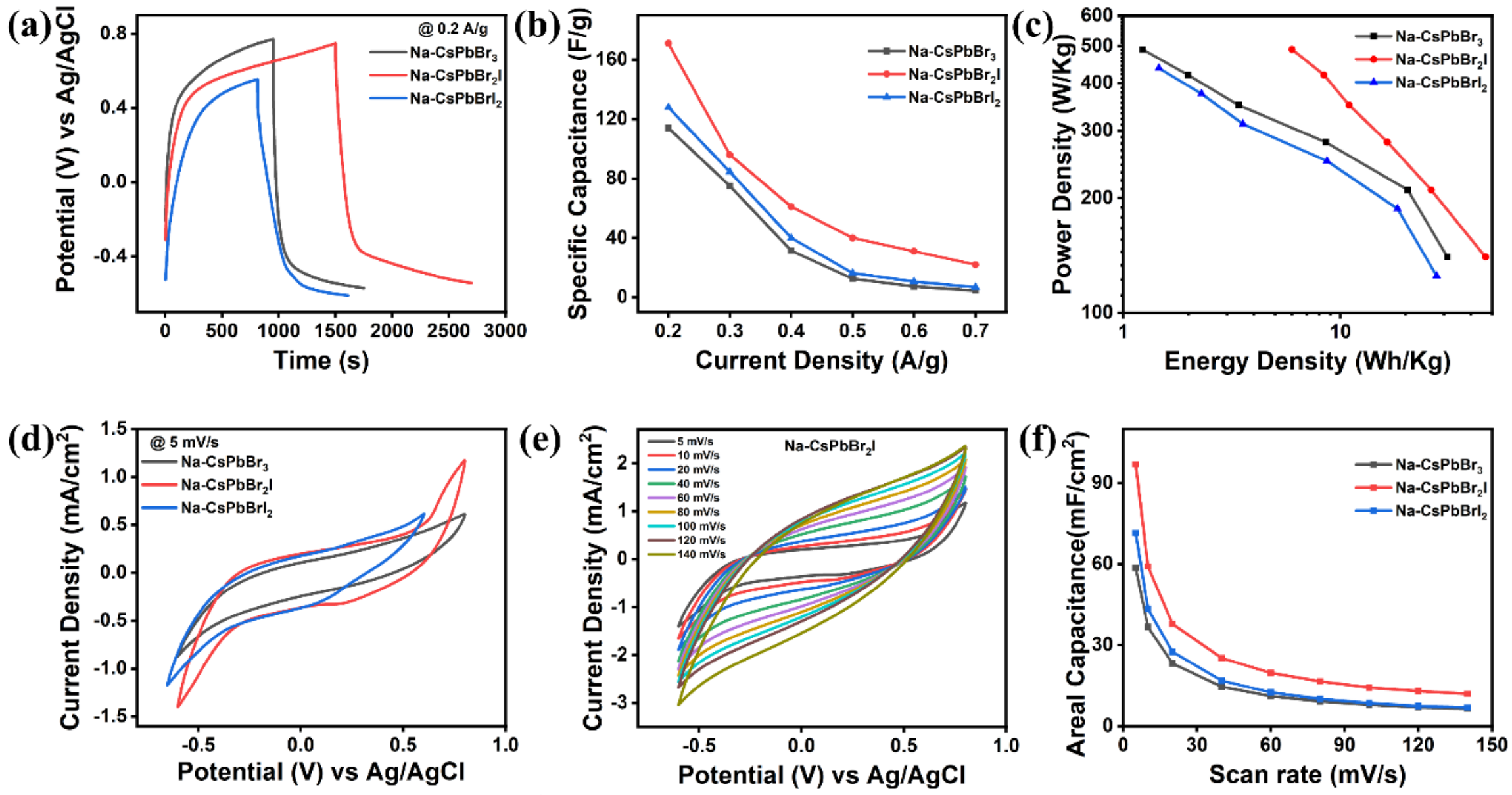

**Figure 4.** Electrochemical response of Na-doped $CsPbBr_{3-x}I_x$ (x = 0, 1, 2) electrodes in 0.1 M tetrabutylammonium tetrafluoroborate (TBTF) in anhydrous dichloromethane, measured in a three-electrode cell with coated graphite as working electrode, Ag/AgCl as reference and platinum as counter electrode. (a) Galvanostatic charge–discharge curves at 0.2 A $g^{-1}$. (b) Specific capacitance against current density from 0.2 to 0.7 A $g^{-1}$, obtained from the discharge branch. (c) Ragone plot of energy against power density on logarithmic axes; points correspond to current densities of 0.2, 0.3, 0.4, 0.5, 0.6 and 0.7 A $g^{-1}$, from lower right to upper left. (d) Cyclic voltammograms at 5 mV $s^{-1}$. (e) Cyclic voltammograms of Na-$CsPbBr_2I$ at scan rates from 5 to 140 mV $s^{-1}$. (f) Areal capacitance against scan rate from 5 to 140 mV $s^{-1}$, obtained by integration of the voltammograms. Active electrode area 1 × 1 $cm^2$; mass loading 1.0 mg $cm^{-2}$.

**Capacitance and rate capability.** Specific capacitances obtained from the discharge branch are plotted against current density in Figure 4(b). At 0.2 A $g^{-1}$ the values are 171 F $g^{-1}$ for Na-$CsPbBr_2I$, 128 F $g^{-1}$ for Na-$CsPbBrI_2$ and 114 F $g^{-1}$ for Na-$CsPbBr_3$, falling to 22, 7 and 5 F $g^{-1}$ respectively at 0.7 A $g^{-1}$. At 0.3 A $g^{-1}$ the corresponding values are 96, 84.5 and 75 F $g^{-1}$. Na-$CsPbBr_2I$ retains the highest capacitance at every current density in this range. Areal capacitances from voltammetry follow the same ordering, at 97, 72 and 59 mF $cm^{-2}$ at 5 mV $s^{-1}$ (Figure 4f). At a loading of 1.0 mg $cm^{-2}$ the areal and specific values of a single measurement are numerically equal, so the two sets differ only through the conditions of measurement, since charge–discharge at 0.3 A $g^{-1}$ corresponds to an effective sweep rate near 3 mV $s^{-1}$ and therefore returns values slightly above those recorded at 5 mV $s^{-1}$. The composition dependence does not follow particle size. Na-$CsPbBrI_2$ has the smallest particles at 11.4 ± 2.8 nm and therefore the largest geometric surface area per unit mass, yet gives a lower capacitance than Na-$CsPbBr_2I$ at every current density, so accessible area is not the controlling variable and the mixed Br:I ratio of 2:1 gives the largest response.

**Behaviour at high sweep rate.** Voltammograms of Na-$CsPbBr_2I$ from 5 to 140 mV $s^{-1}$ are shown in Figure 4(e). The curves remain closed and free of redox peaks across the whole range, but they tilt progressively and the current does not grow in proportion to the sweep rate: the value at 0.5 V rises from 0.40 to 1.71 mA $cm^{-2}$, a factor of 4.2, against the factor of 28 that a purely capacitive response would give. Areal capacitance falls correspondingly, from 97 to 12 mF $cm^{-2}$ over the same range, retaining 12% of its low-rate value (Figure 4f). Charge storage in these electrodes is therefore limited by ion transport rather than by the interfacial process once the sweep rate exceeds a few tens of millivolts per second. Tetrabutylammonium and tetrafluoroborate ions in dichloromethane are larger and less mobile than aqueous ions, so at high sweep rates only the outer surface of the porous electrode is reached within the time available, while at low rates ions penetrate the mesopore network.[48] The same ordering holds throughout, with Na-$CsPbBr_2I$ above the other two compositions at every scan rate.

**Energy and power density.** The Ragone plot in Figure 4(c) pairs the two quantities across the current-density range. Na-$CsPbBr_2I$ delivers 47 W h $kg^{-1}$ at 140 W $kg^{-1}$ and 26 W h $kg^{-1}$ at 210 W $kg^{-1}$, against 31 and 20 W h $kg^{-1}$ at the same two power densities for Na-$CsPbBr_3$ and 28 and 18 W h $kg^{-1}$ for Na-$CsPbBrI_2$. The Na-$CsPbBr_2I$ curve lies above and to the right of the other two throughout, so its advantage is maintained across the range rather than confined to a single operating point, and the separation between the three widens as power density falls. These values were obtained in the three-electrode configuration, in which the response of the perovskite electrode is isolated against a reference. Capacitance, energy density and power density measured this way lie above those returned by a symmetric two-electrode cell of the same materials, so the figures reported here are electrode-level rather than device-level. Absolute values in this system are set as much by the electrolyte as by the electrode: reported capacitances for halide perovskite electrodes span two orders of magnitude, from 528 mF $g^{-1}$ for $CsPbBr_3$ nanocrystals in a quaternary ammonium salt in dichloromethane [10] to 121 F $g^{-1}$ for $CsPbBr_3$ in aqueous electrolyte measured by voltammetry at the same 5 mV $s^{-1}$ used here [9] and 886 F $g^{-1}$ for $CsPb_2Br_5$. [11,15] Comparison here is therefore made against the undoped materials measured in the same cell and the same electrolyte.

### 3.5 Interfacial and Ion-Transport Kinetics

**Impedance response.** Impedance spectra were recorded from 0.01 Hz to 100 kHz. The Nyquist plots in Figure 5(a) show a depressed semicircle at high frequency followed by a steeply rising branch at low frequency, the response of a porous electrode in which a charge-transfer process at the interface is followed by finite ion diffusion within the pore network.[44,49,50] Charge-transfer resistances taken from the width of the high-frequency semicircle are compared in Figure 5(b): 175 Ω for Na-$CsPbBr_2I$, 242 Ω for Na-$CsPbBrI_2$ and 295 Ω for Na-$CsPbBr_3$. The composition with the lowest interfacial resistance is the one with the highest capacitance, which places the origin of the capacitance difference at the interface rather than in the accessible surface area, since Na-$CsPbBrI_2$ has the smallest particles and therefore the largest geometric area yet a charge-transfer resistance 67 Ω higher than that of Na-$CsPbBr_2I$. Spectra recorded at DC bias values between 0 and 1.0 V gave charge-transfer resistances constant to within the measurement uncertainty for all three compositions, so no potential-dependent faradaic process is switched on within the operating window (Figure S9).

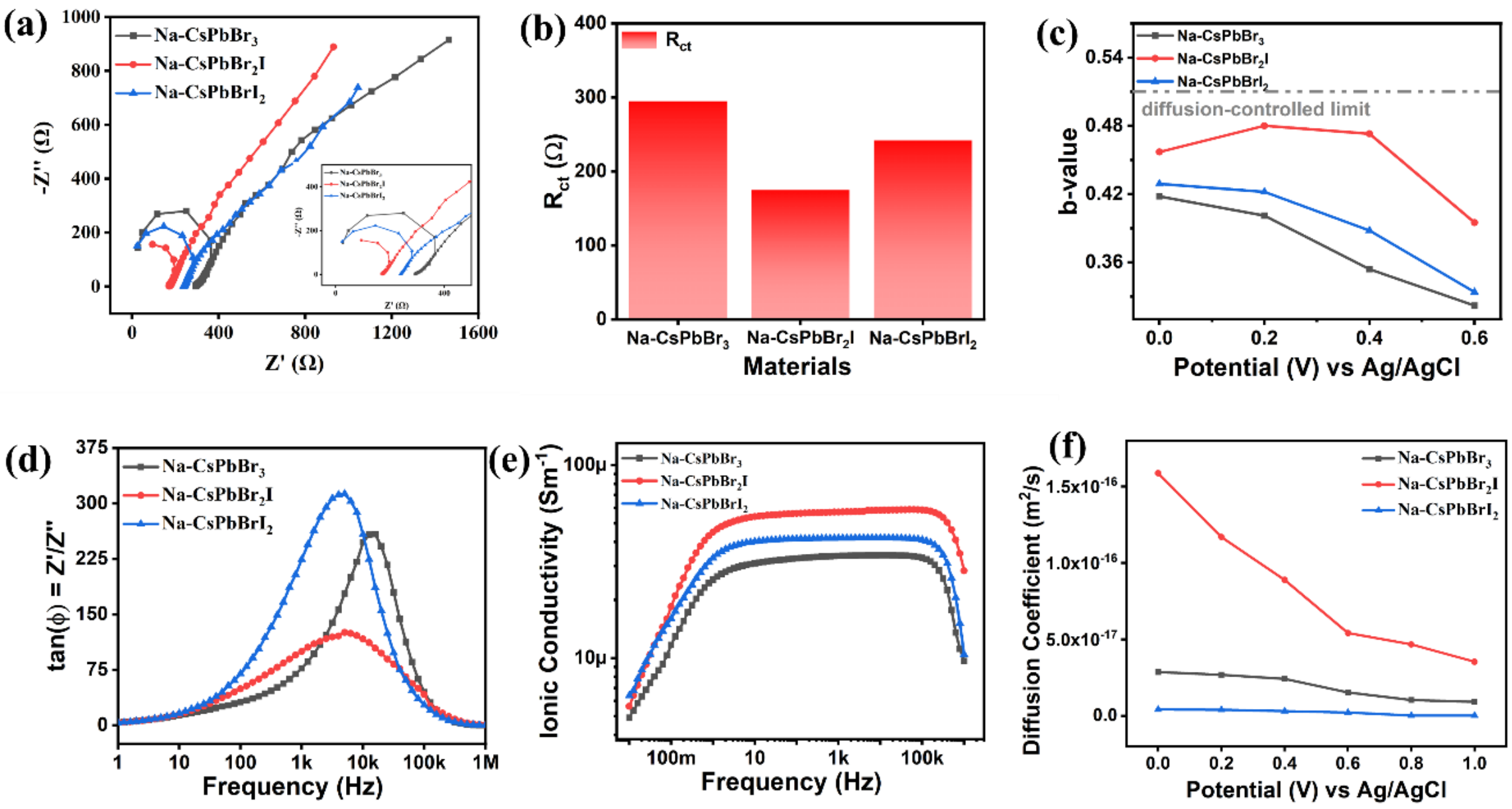

**Figure 5.** Interfacial and ion-transport kinetics of Na-doped $CsPbBr_{3-x}I_x$ electrodes in 0.1 M tetrabutylammonium tetrafluoroborate (TBTF) in anhydrous dichloromethane, three-electrode cell against Ag/AgCl. (a) Nyquist plots recorded from 0.01 Hz to 100 kHz at open-circuit potential, with the high-frequency region enlarged in the inset. (b) Charge-transfer resistance for each composition. (c) Power-law exponent $b$ of the relation $i = av^b$ obtained at fixed potentials between 0 and 0.6 V from voltammograms recorded at 5 to 80 mV $s^{-1}$; the dashed line marks $b = 0.5$, the limit for diffusion-controlled storage. (d) Loss tangent against frequency. (e) Frequency-dependent conductivity of the composite electrode film. (f) Ion diffusion coefficient against applied potential, obtained from the low-frequency impedance.

**Kinetics of charge storage.** The dependence of voltammetric current on sweep rate distinguishes charge stored at the surface from charge stored through a diffusion-limited process. Writing the current at a fixed potential as $i = av^b$, an exponent of unity corresponds to surface-controlled storage and an exponent of one half to diffusion control.[51–53] Values obtained from the voltammograms of Figure 4(e) and its equivalents are plotted against potential in Figure 5(c), with the log–log fit for Na-$CsPbBr_2I$ at 0.2 V shown in Figure S10. Na-$CsPbBr_2I$ gives the highest exponent at every potential, averaging 0.451 between 0 and 0.6 V, against 0.395 for Na-$CsPbBrI_2$ and 0.374 for Na-$CsPbBr_3$. The exponent decreases with increasing potential for all three compositions, by 0.06 for Na-$CsPbBr_2I$ and by 0.10 for the other two across the same interval.

Every value lies at or below 0.5, the limit for diffusion-controlled storage, so neither limiting case describes these electrodes and the separation of the current into capacitive and diffusion-controlled components is not applicable here.[54] An exponent below 0.5 arises when ion transport through the pore network, rather than the interfacial process itself, sets the rate at which charge can be stored. That reading is consistent with the fall of areal capacitance from 97 to 12 mF $cm^{-2}$ between 5 and 140 mV $s^{-1}$ (Figure 4f), with the charge-transfer resistances of 175 to 295 Ω, and with the decrease of the exponent as the electrode is polarised and the pore network becomes progressively depleted.

**Relaxation and ion transport.** The loss tangent in Figure 5(d) passes through a maximum whose position gives the characteristic relaxation frequency of the electrode, from which the relaxation time follows as $\tau = 1/(2\pi f_{max})$. Maxima occur at 15.8 kHz for Na-$CsPbBr_3$, 5.0 kHz for Na-$CsPbBr_2I$ and 5.0 kHz for Na-$CsPbBrI_2$, giving relaxation times of 63, 200 and 200 μs. Frequency-dependent conductivity is shown in Figure 5(e), where all three materials show the pattern expected of an ionic conductor: a low-frequency dispersion where charge accumulates at the blocking interface, a plateau across the mid-frequency range that gives the direct-current conductivity, and a fall at the highest frequencies. Plateau values are 60.3 μS $m^{-1}$ for Na-$CsPbBr_2I$, 41.7 μS $m^{-1}$ for Na-$CsPbBrI_2$ and 33.7 μS $m^{-1}$ for Na-$CsPbBr_3$, ordered as the charge-transfer resistances are; the conductivity here refers to the composite electrode film of thickness 11.3 μm (Figure S11) rather than to the bulk electrolyte. Diffusion coefficients obtained from the low-frequency impedance by the Bandara–Mellander treatment[33] are plotted against applied potential in Figure 5(f). Na-$CsPbBr_2I$ gives $1.6 \times 10^{-16}$ $m^2$ $s^{-1}$ at 0 V, falling to $3.5 \times 10^{-17}$ $m^2$ $s^{-1}$ at 1.0 V. Na-$CsPbBr_3$ gives $2.8 \times 10^{-17}$ $m^2$ $s^{-1}$ at 0 V falling to $9.1 \times 10^{-18}$ $m^2$ $s^{-1}$, and Na-$CsPbBrI_2$ remains below $5 \times 10^{-18}$ $m^2$ $s^{-1}$ across the whole range. The diffusion coefficient of Na-$CsPbBr_2I$ is therefore between five and six times that of Na-$CsPbBr_3$ at zero bias and around thirty times that of Na-$CsPbBrI_2$. The decrease with increasing bias is common to all three: as the electrode is polarised, ions accumulate in the double layer at the pore walls, the local concentration gradient driving further transport falls, and the effective diffusion coefficient drops. Figures 6(c) and 6(f) therefore describe the same behaviour from two independent measurements, one voltammetric and one impedance-based, with both quantities falling as the electrode is polarised. Four measurements rank the three compositions identically. Na-$CsPbBr_2I$ has the highest capacitance, the lowest charge-transfer resistance, the highest ion diffusion coefficient and the highest power-law exponent, and it is the composition in which ion transport limits charge storage least severely. These are properties of the electrolyte-filled pore network rather than of the perovskite lattice, and Section 3.6 turns to the behaviour of the lattice itself under prolonged cycling.

### 3.6 Cycling Stability and Suppression of the Capacitance-Rise Signature

Figure 6 compares the undoped and sodium-containing series under identical conditions — the same electrolyte, the same three-electrode configuration, and the same 2500-cycle protocol — and this comparison is the measurement that tests the design hypothesis of Section 3.1. Specific capacitance at 0.3 A $g^{-1}$ rises on sodium incorporation for every composition, from 42 to 75 F $g^{-1}$ for $CsPbBr_3$, from 63 to 96 F $g^{-1}$ for $CsPbBr_2I$ and from 56 to 84.5 F $g^{-1}$ for $CsPbBrI_2$, corresponding to gains of 79, 52 and 51% (Figure 6a).

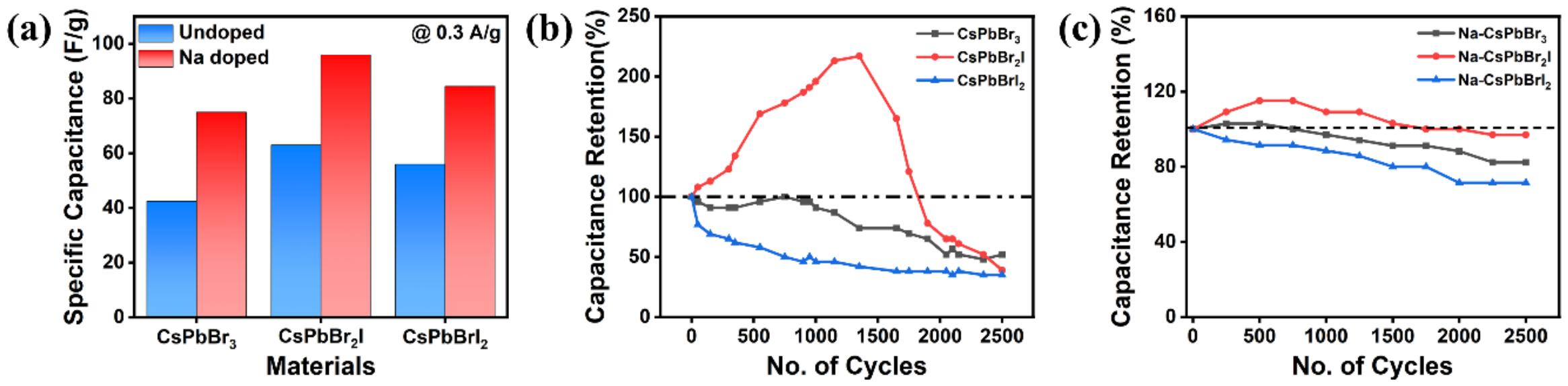

**Figure 6.** Capacitance and cycling stability of undoped and Na-doped $CsPbBr_{3-x}I_x$ (x = 0, 1, 2) electrodes. (a) Specific capacitance at 0.3 A $g^{-1}$ for the undoped and sodium-containing materials. (b) Capacitance retention of the undoped materials over 2500 galvanostatic charge–discharge cycles at 0.5 A $g^{-1}$, normalised to the first-cycle value. (c) The same measurement for the sodium-containing materials; note the different vertical scale, and the dashed line in both panels marks 100% retention.

The retention curves of the undoped materials in Figure 6(b) reproduce the behaviour reported previously.[22] $CsPbBr_3$ holds close to its initial capacitance for the first 750 cycles and then declines steadily to 52% at cycle 2500, and $CsPbBrI_2$ declines from the outset to 35%. $CsPbBr_2I$ follows neither pattern: its capacitance climbs continuously to 218% of the first-cycle value by cycle 1378, falls sharply between cycles 1500 and 2000, and reaches 39% at cycle 2500. The rise and the collapse are two stages of one process, because as iodide is expelled from the lattice under the applied field the pore volume increases and the accessible surface area with it, so capacitance climbs until enough of the framework has been removed that the remaining structure can no longer support the interface. Loss of the I 3d photoemission signal and the appearance of CsI and $PbBr_2$ in the diffraction pattern of the cycled electrode place that process on the halide sublattice rather than on the electrolyte or the binder.[22] Against this, the sodium-containing series in Figure 6(c) behaves differently in degree rather than in kind. Na-$CsPbBr_2I$ rises to a maximum of 115% between cycles 500 and 750, holds within a few percent of that value to cycle 1250, and then declines gradually to 97% at cycle 2500, with no collapse. Na-$CsPbBr_3$ reaches 103% between cycles 250 and 500 before falling to 82%, and Na-$CsPbBrI_2$ decays monotonically to 71%. For the $CsPbBr_2I$ composition the maximum of the rise therefore falls from 218 to 115% of the first-cycle capacitance and moves earlier by roughly 750 cycles, from cycle 1378 to cycle 600, while retention at 2500 cycles rises from 39 to 97%. Expressed as the excess above the first-cycle value, the rise is reduced by a factor of 7.9. The residual 15% matters as much as the reduction. Reading the rise as the signature of field-driven halide expulsion, a maximum of 115% rather than 100% means that redistribution still occurs over the first several hundred cycles, at a rate low enough that it does not proceed to structural collapse within 2500 cycles. Sodium incorporation therefore suppresses the process rather than preventing it, and the measurement places an upper bound on the extent of lattice reorganisation rather than establishing that none occurs. Retention improves for all three compositions, from 52 to 82%, from 39 to 97% and from 35 to 71%, and the largest gain belongs to the composition that showed the rise-and-collapse behaviour, which is what would be expected if sodium acts on halide transport rather than on a degradation route common to all three. Three measurements made by different techniques point the same way. Iodine is retained in the cycled Na-$CsPbBr_2I$ electrode at unchanged binding energy where the undoped electrode lost its I 3d signal entirely (Section 3.3), the Br:I coverage ratio measured by elemental mapping is 2.30 after cycling against 2.25 before, and the capacitance rise that accompanied iodine expulsion in the undoped material is reduced by a factor of 7.9. The chemical-state and elemental measurements reported here describe the electrode surface, so the long-range crystallographic state of the cycled material is not established from them. Retention of 97% over 2500 cycles sits within the range reported for halide perovskite supercapacitor electrodes, which spans 73% over 5000 cycles for $CsPbBr_3$ in aqueous electrolyte,[9] 76% over 5000 cycles for $CsPb_2Br_5$ and 89% over 5000 cycles for a $CsPbBr_3$ thin-film device, [11] and 90% over 10,000 cycle[11] and 90% over 10,000 cycles for $CsPbBr_3$ nanocrystals in a quaternary ammonium salt in dichloromethane.[10] Cycle counts, electrolytes and cell configurations differ across those studies, which limits direct comparison. The comparison that carries the argument here is the one against the undoped materials in the same cell and the same electrolyte, where the only variable is the presence of sodium.

## 4. Conclusion

Sodium was incorporated into $CsPbBr_3$, $CsPbBr_2I$ and $CsPbBrI_2$ nanocrystals during ligand-assisted reprecipitation at a Na/Pb precursor ratio of 1.25:1.00, and the doped and undoped materials were compared as supercapacitor electrodes in 0.1 M tetrabutylammonium tetrafluoroborate in anhydrous dichloromethane under an identical protocol. The nanocrystals are single-crystalline orthorhombic perovskite with cuboidal habit and mean edge lengths of $19.0 \pm 4.6$, $18.2 \pm 5.2$ and $11.4 \pm 2.8$ nm for the three compositions. Sodium incorporation contracts the lattice, reducing the pseudocubic parameter from $5.908 \pm 0.029$ to $5.851 \pm 0.019$ Å after correction for specimen displacement, and the parameter increases monotonically with iodide content across both series. Specific capacitance at 0.3 A $g^{-1}$ rises on sodium incorporation for every composition, from 42 to 75 F $g^{-1}$, from 63 to 96 F $g^{-1}$ and from 56 to 84.5 F $g^{-1}$, and the improvement holds across the current-density range from 0.2 to 0.7 A $g^{-1}$. Na-$CsPbBr_2I$ gives the lowest charge-transfer resistance of the three doped compositions at 175 Ω, against 242 Ω for Na-$CsPbBrI_2$ and 295 Ω for Na-$CsPbBr_3$, the highest ion diffusion coefficient at $1.6 \times 10^{-16}$ $m^2$ $s^{-1}$ at zero bias, and the highest power-law exponent for the dependence of current on sweep rate, averaging 0.451 between 0 and 0.6 V. Every exponent measured lies at or below 0.5, so charge storage in these electrodes is limited by ion transport through the pore network rather than by the interfacial process. The cycling comparison establishes the main result. Undoped $CsPbBr_2I$ rises to 218% of its first-cycle capacitance by cycle 1378 and then falls to 39% by cycle 2500. In the sodium-containing analogue the rise reaches 115% at cycle 600 and no collapse follows, with 97% of the first-cycle capacitance retained at cycle 2500. Expressed as the excess above the first-cycle value, the rise is reduced by a factor of 7.9 and its maximum moves earlier by 750 cycles. Retention at 2500 cycles improves for every composition, from 52 to 82%, from 39 to 97% and from 35 to 71%. Chemical-state measurements on the cycled electrodes are consistent with this. Both I 3d components are present after 2500 cycles at unchanged binding energy, and the Br:I ratio measured by elemental mapping is 2.30 after cycling against 2.25 before, whereas the I 3d signal was absent from the undoped electrode cycled under the same protocol.[22] Lead remains divalent, with no metallic component detected, and sodium is retained in the cycled electrode. These measurements establish that sodium incorporation suppresses, without eliminating, the capacitance rise that precedes structural collapse in undoped $CsPbBr_2I$, and that iodine is retained in the cycled electrode. They do not establish the lattice site occupied by sodium, the crystallographic state of the cycled material, or the magnitude of the halide migration barrier.

# *Supporting Information*

## Sodium Incorporation in $CsPbBr_{3-x}I_x$ Nanocrystal Electrodes: Lattice Contraction and the Suppression of Field-Driven Iodine Expulsion

*Arun Kumar*[†], *Monojit Bag*[†, ‡, *].

†Advanced Research in Electrochemical Impedance Spectroscopy Laboratory, Indian Institute of Technology Roorkee, Roorkee 247667, India

‡Centre for Nanotechnology, Indian Institute of Technology Roorkee, Roorkee 247667, India

# Calculations

**Areal ($C_A$)** and **specific ($C_S$) capacitance** calculated from CV measurements using these equations.

$$C_A = \frac{\int(IdV)}{V \times s \times A} \tag{S1}$$

$$C_s = \frac{\int(IdV)}{V \times s \times m} \tag{S2}$$

**Areal ($C_A$)** and **specific ($C_S$) capacitance** calculated from GCD measurements using these equations.

$$C_A = \frac{I\int(Vdt)}{m \times V^2} \tag{S3}$$

$$C_s = \frac{I\int(Vdt)}{A \times V^2} \tag{S4}$$

Energy density and power density calculated as

$$E = \frac{1}{7.2} \times C_s(V)^2 \tag{S5}$$

$$P = \frac{E}{t} \tag{S6}$$

where $\int$ (IdV) is the active area under CV spectra, V is potential window (V), s is the scan rate (mV $s^{-1}$), A is the active electrode area, m is mass of the active material (g) and t is the galvanostatic discharging time.

**Bandara-Mellander (B-M) Formalism**

The dielectric loss tangent can be calculated directly from electrochemical impedance data using this following relation:

$$\tan(\phi) = \frac{Z'}{Z''} \tag{S7}$$

The Bandara and Mellander (B-M) model, which considers the effect of space charge polarization in its estimation of ionic diffusion coefficient, is used for that purpose. Using the B-M model, the relaxation time for the space charge and all other necessary parameters are calculated using either the loss tangent plot or the boundary condition of the dielectric function. This has been widely used by various research groups in recent investigations of the metal halide perovskite (MHP)-based optoelectronic device. Using this formulation, both real ($\epsilon'$) and imaginary ($\epsilon''$) parts of the complex dielectric constant, as well as the dielectric loss: tangent, can be represented as:

$$\epsilon' = \epsilon'_\infty \left(1 + \frac{\delta}{1 + (\omega\tau_1\delta)^2}\right) \tag{S8}$$

$$\epsilon'' = \epsilon'_\infty \left(1 + \frac{\omega\tau_1\delta^2}{1 + (\omega\tau_1\delta)^2}\right) \tag{S9}$$

$$\tan(\phi) = \frac{\epsilon''}{\epsilon'} = \frac{\omega\tau_1\delta}{1 + \omega^2\tau_1^2\delta} \tag{S10}$$

Here, $\tau_1$ represents the recombination time, which is related to the macroscopic relaxation time ($\tau_2$) through the following equation:

$$\tau_2 = \tau_1\sqrt{\delta} \tag{S11}$$

The point of maximum of the loss tangent (tan(ϕ)) function is directly proportional to $\tau_2$, whereas the parameter δ is derived from the loss tangent maximum. Hence, the loss tangent spectrum obtained experimentally can be described through the tan(ϕ) expression as described above.

Finally, by means of B-M method, the macroscopic ion diffusion coefficient ($D_i$) is expressed as:

$$D_i = \frac{L^2}{\tau_2 \delta^2} \tag{S12}$$

where $L$ denotes the thickness of the electrode.

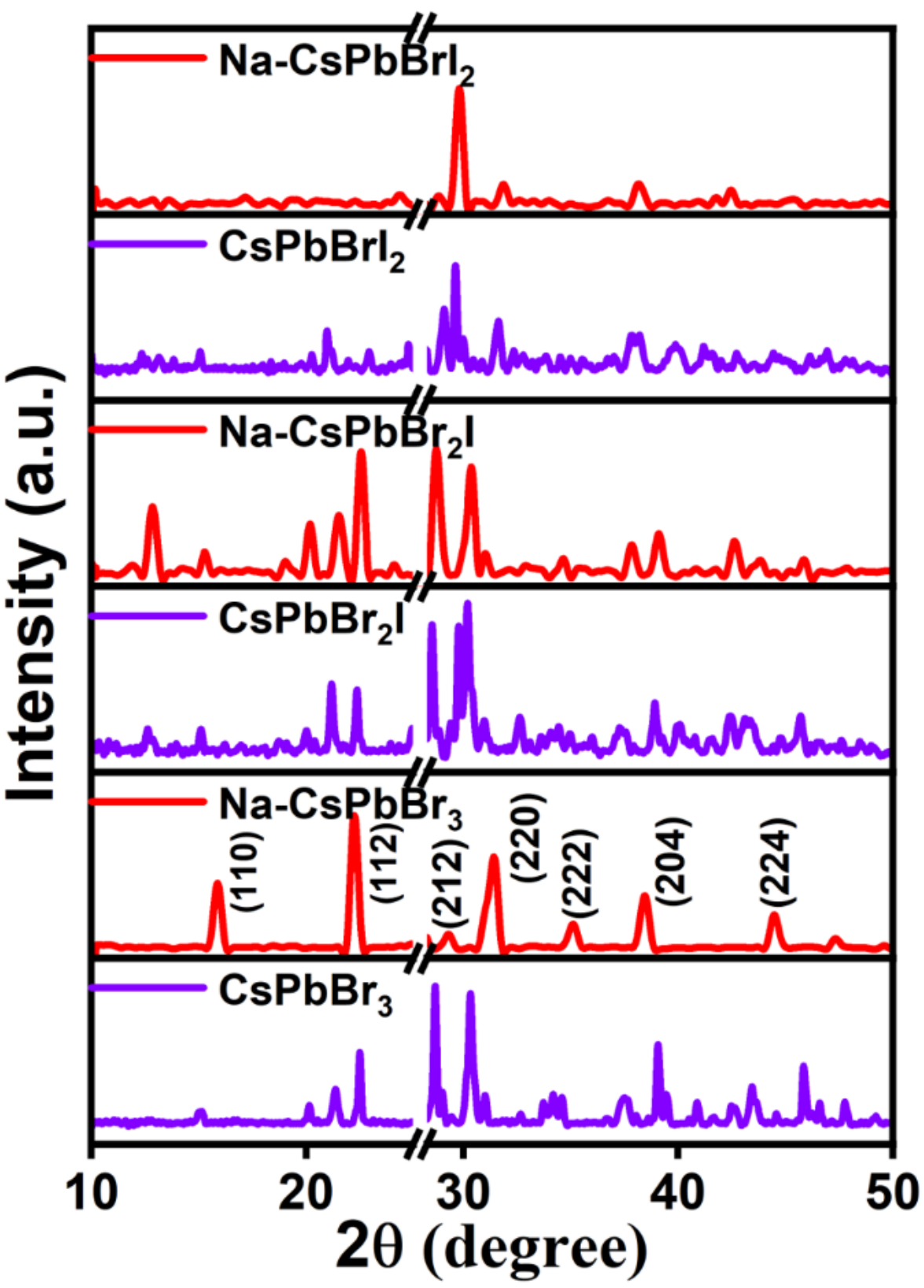

**Figure S1.** XRD compared of undoped and Na doped $CsPbBr_{3-x}I_x$ NCs from 10 to 50 degree.

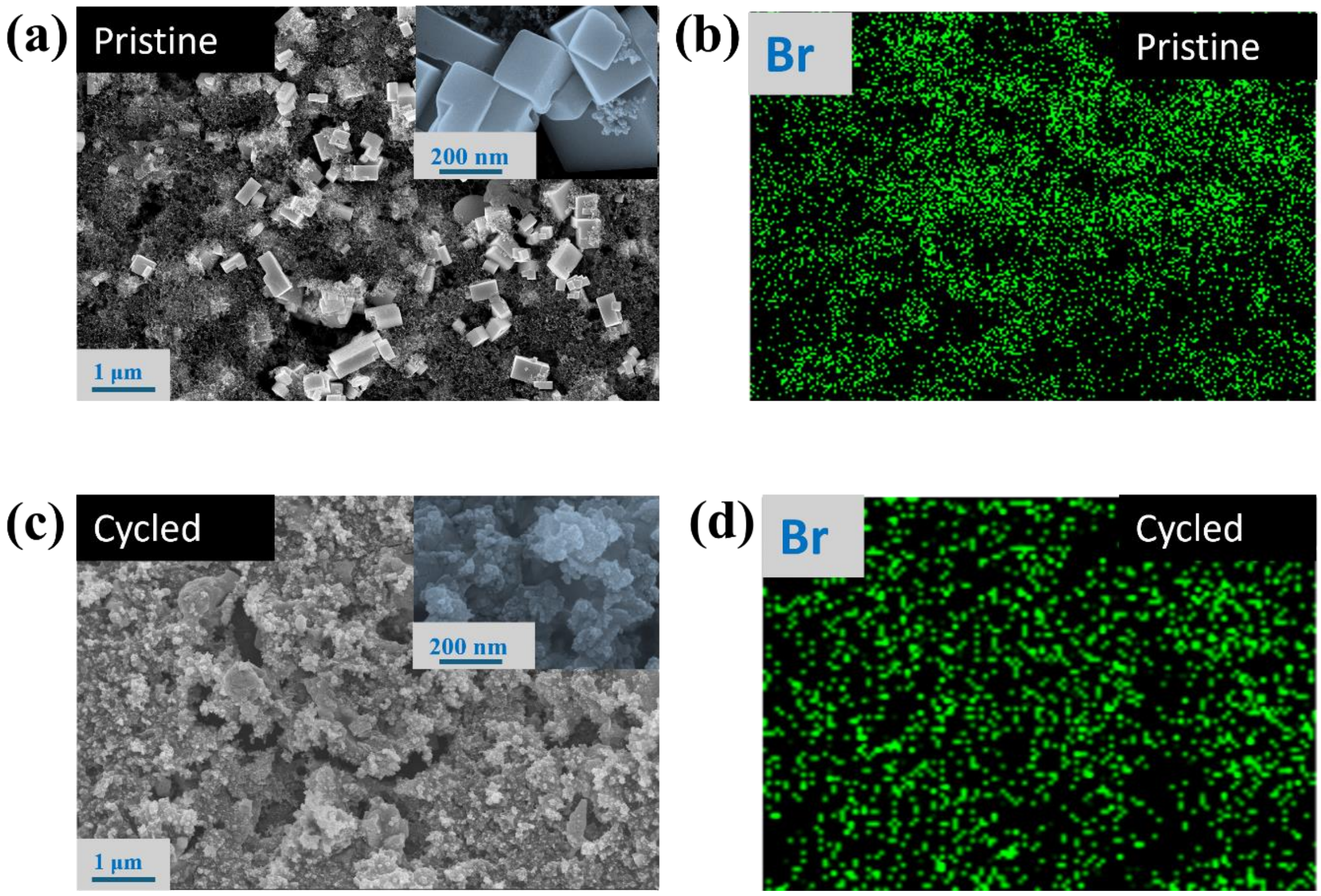

**Figure S2.** FE-SEM images and EDS elemental mapping spectra for the (a, b) pristine and (c, d) cycled electrodes of Na-$CsPbBr_3$.

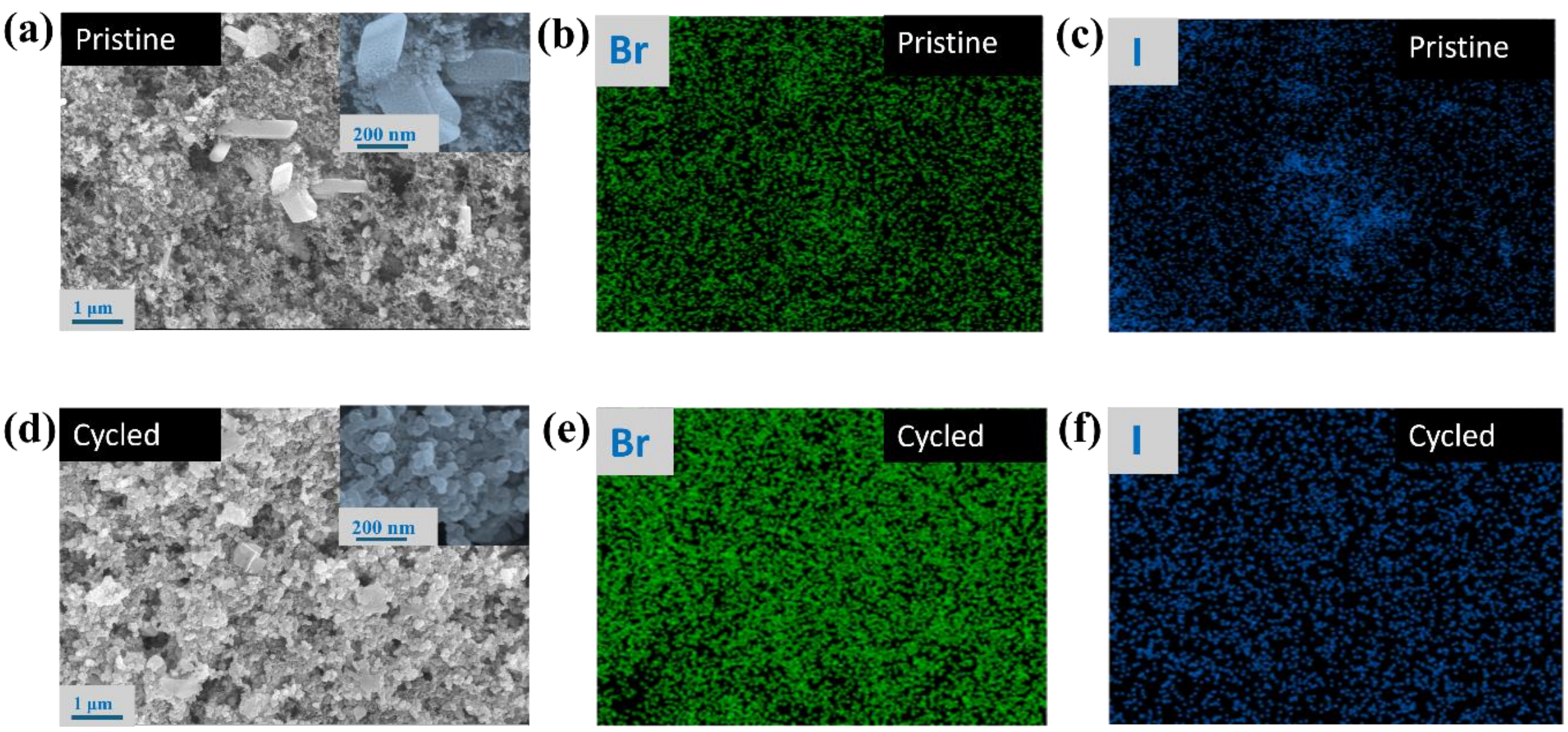

**Figure S3.** FE-SEM images and EDS elemental mapping spectra for the (a-c) pristine and (d-f) characterized electrodes of Na-$CsPbBrI_2$.

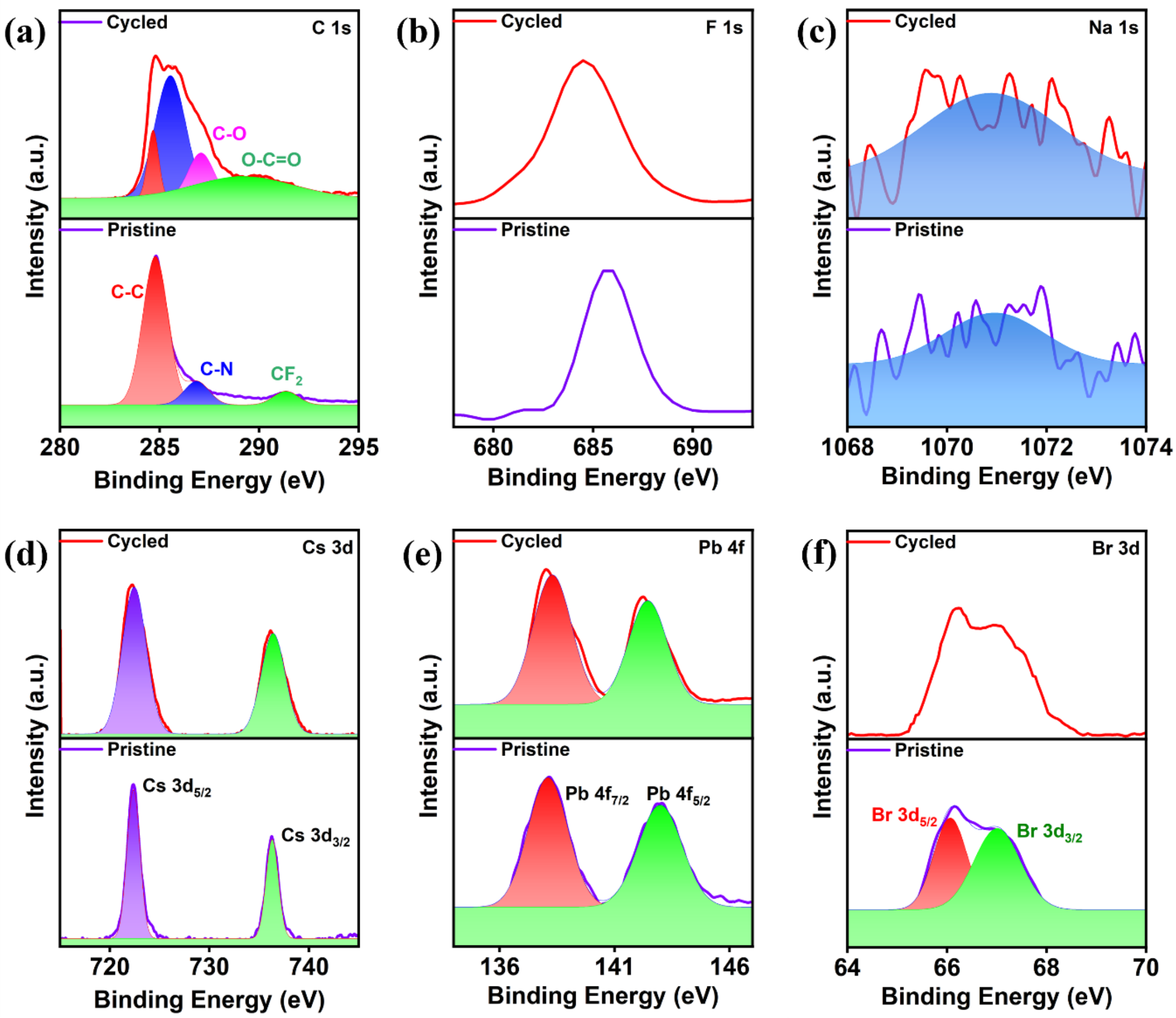

**Figure S4.** XPS analysis of Na-doped $CsPbBr_3$ perovskite before and after electrochemical characterization: high-resolution spectra of (a) C 1s, (b) F 1s, (c) Na 1s, (d) Cs 3d, (e) Pb 4f, and (f) Br 3d.

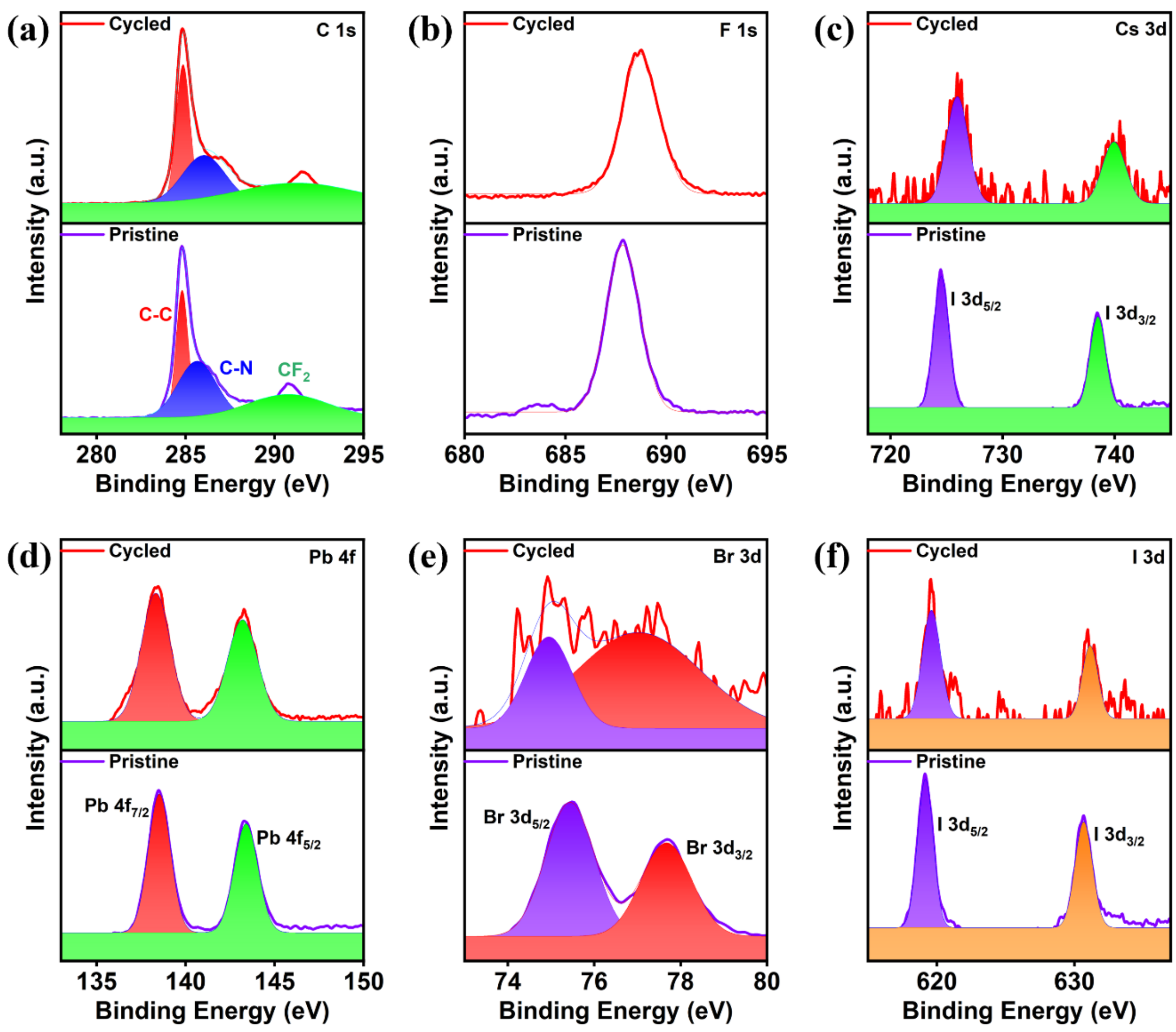

**Figure S5.** XPS analysis of Na-doped $CsPbBrI_2$ perovskite before and after electrochemical characterization: (a) C 1s, (b) F 1s, (c) Cs 3d, (d) Pb 4f, (e) Br 3d, and (f) I 3d.

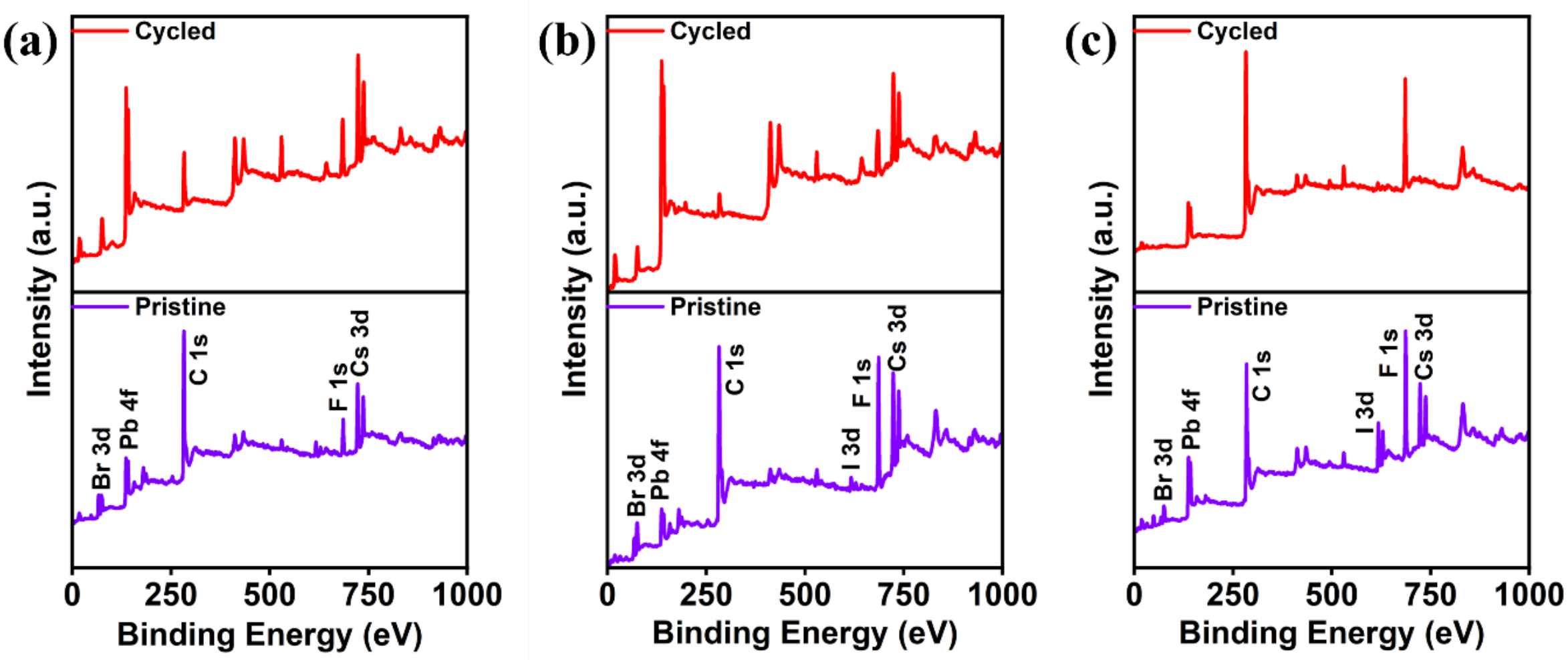

**Figure S6.** XPS analysis survey spectra of Na-doped (a) $CsPbBr_3$, (b) $CsPbBr_2I$ and (c) $CsPbBrI_2$ perovskite before and after electrochemical characterization.

| | Na-$CsPbBr_3$ (Atomic %) | | Na-$CsPbBr_2I$ (Atomic %) | | Na-$CsPbBrI_2$ (Atomic %) | |
|---|---|---|---|---|---|---|
| **Element** | **Pristine** | **Cycled** | **Pristine** | **Cycled** | **Pristine** | **Cycled** |
| C | 71.71 | 77.12 | 86.54 | 90.46 | 91.57 | 80.70 |
| F | 16.49 | 15.05 | 2.64 | 3.08 | 4.66 | 15.01 |
| Na | 0.71 | 0.64 | 0.73 | 0.68 | 0.68 | 0.68 |
| Br | 3.18 | 2.07 | 4.06 | 2.47 | 0.29 | 0.58 |
| I | ___ | ___ | 1.09 | 0.83 | 0.99 | 0.43 |
| Cs | 4.04 | 2.34 | 3.50 | 1.75 | 0.73 | 1.62 |
| Pb | 3.87 | 2.78 | 1.44 | 0.73 | 1.08 | 0.89 |
| **Total** | **100.00** | **100.00** | **100.00** | **100.00** | **100.00** | **100.00** |

**Table S1:** EDS elemental composition in atomic % for all three compositions, pristine and cycled.

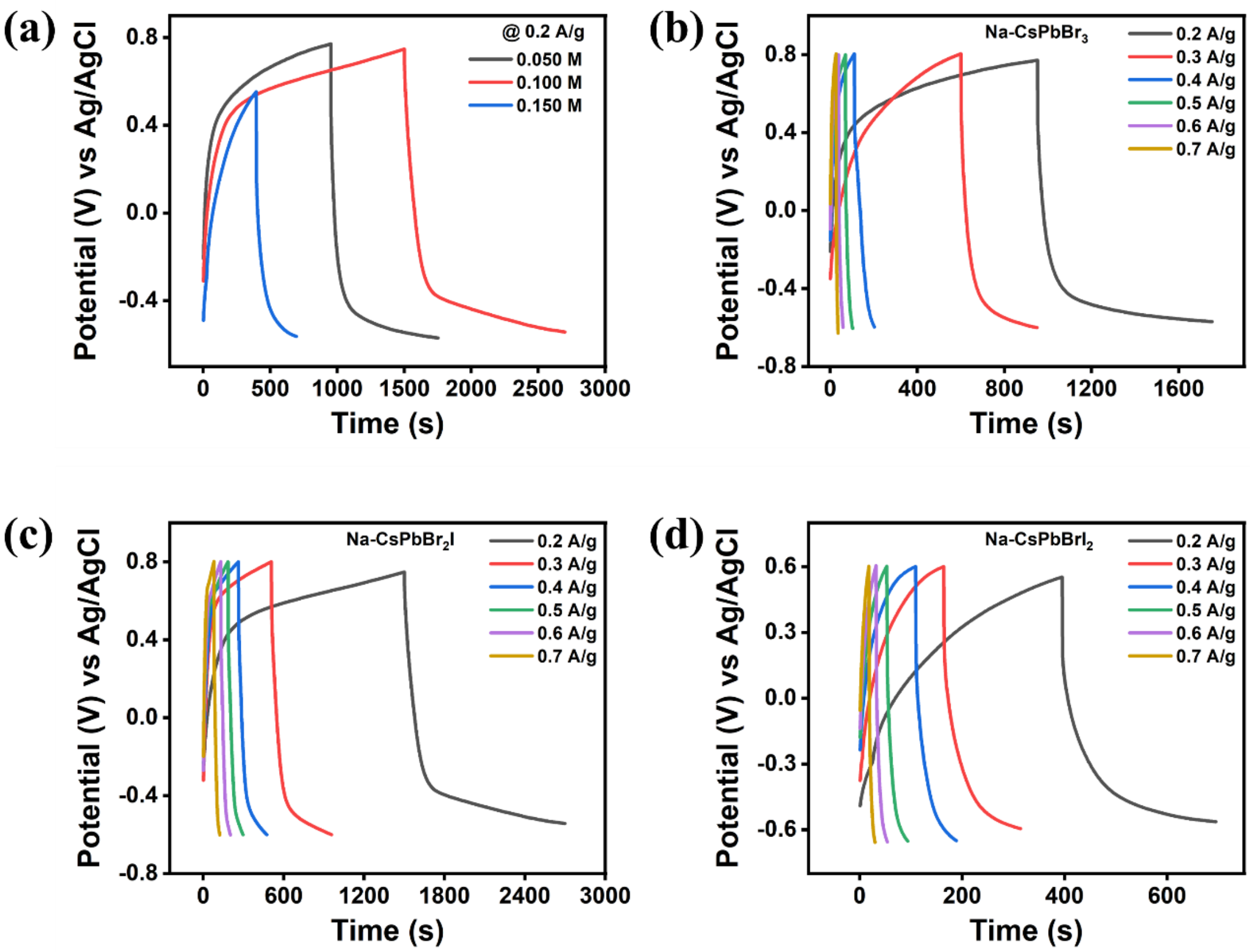

**Figure S7.** Electrochemical storage performances of trace Na-$CsPbBr_{3-x}I_x$ electrodes. (a) GCD curves obtained at a current density of 0.2 A/g. GCD profile of (b) Na-$CsPbBr_3$, (c) Na-$CsPbBr_2I$, (d) and $NaCsPbBrI_2$ at current densities from 0.2 to 0.7 A/g.

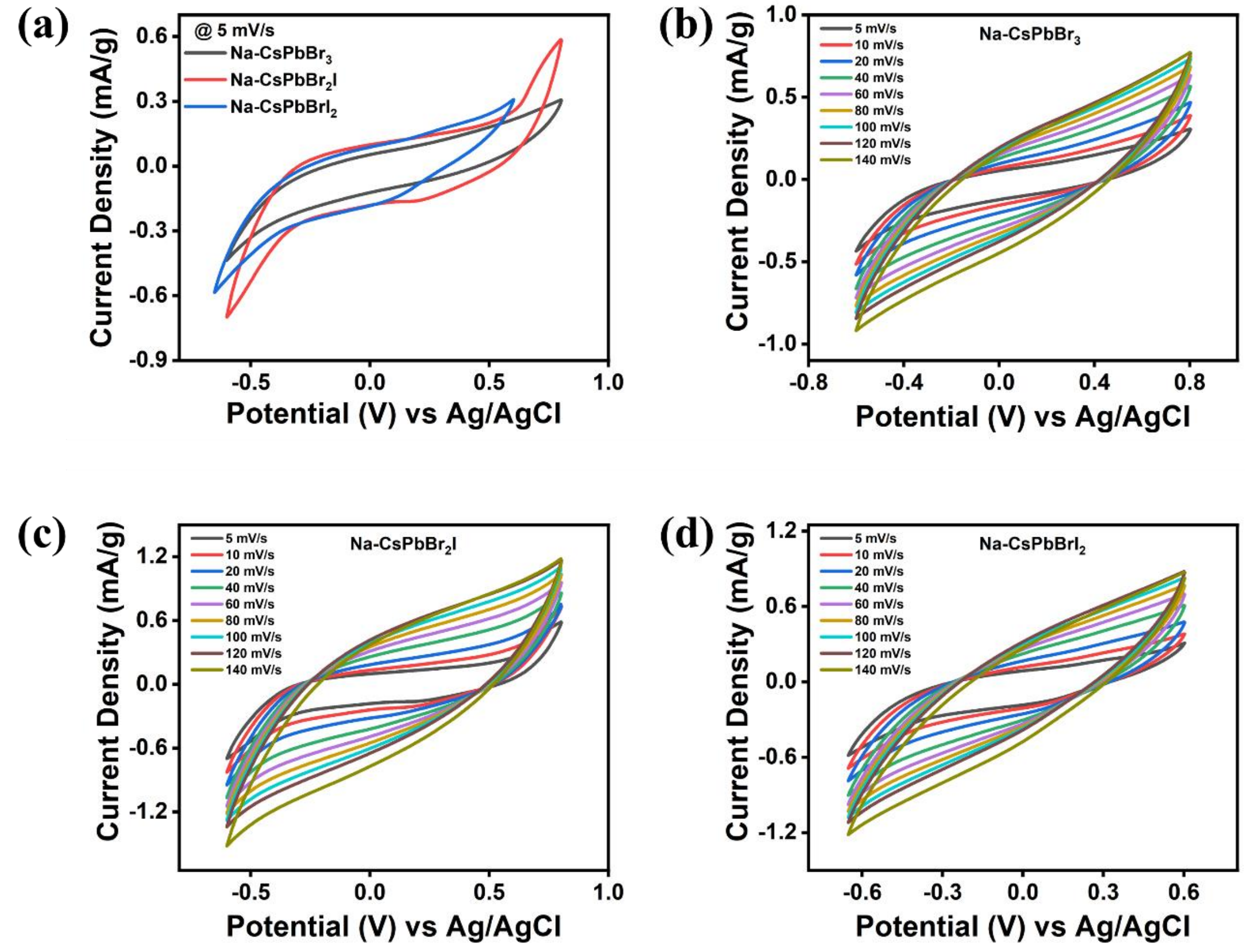

**Figure S8.** Electrochemical storage performances of trace Na-$CsPbBr_{3-x}I_x$ electrodes. (a) CV curves obtained at a scan rate of 5 mV/s. CV profile of (b) Na-$CsPbBr_3$, (c) Na-$CsPbBr_2I$, (d) and $NaCsPbBrI_2$ at scan rate from 5 to 140 mV/s.

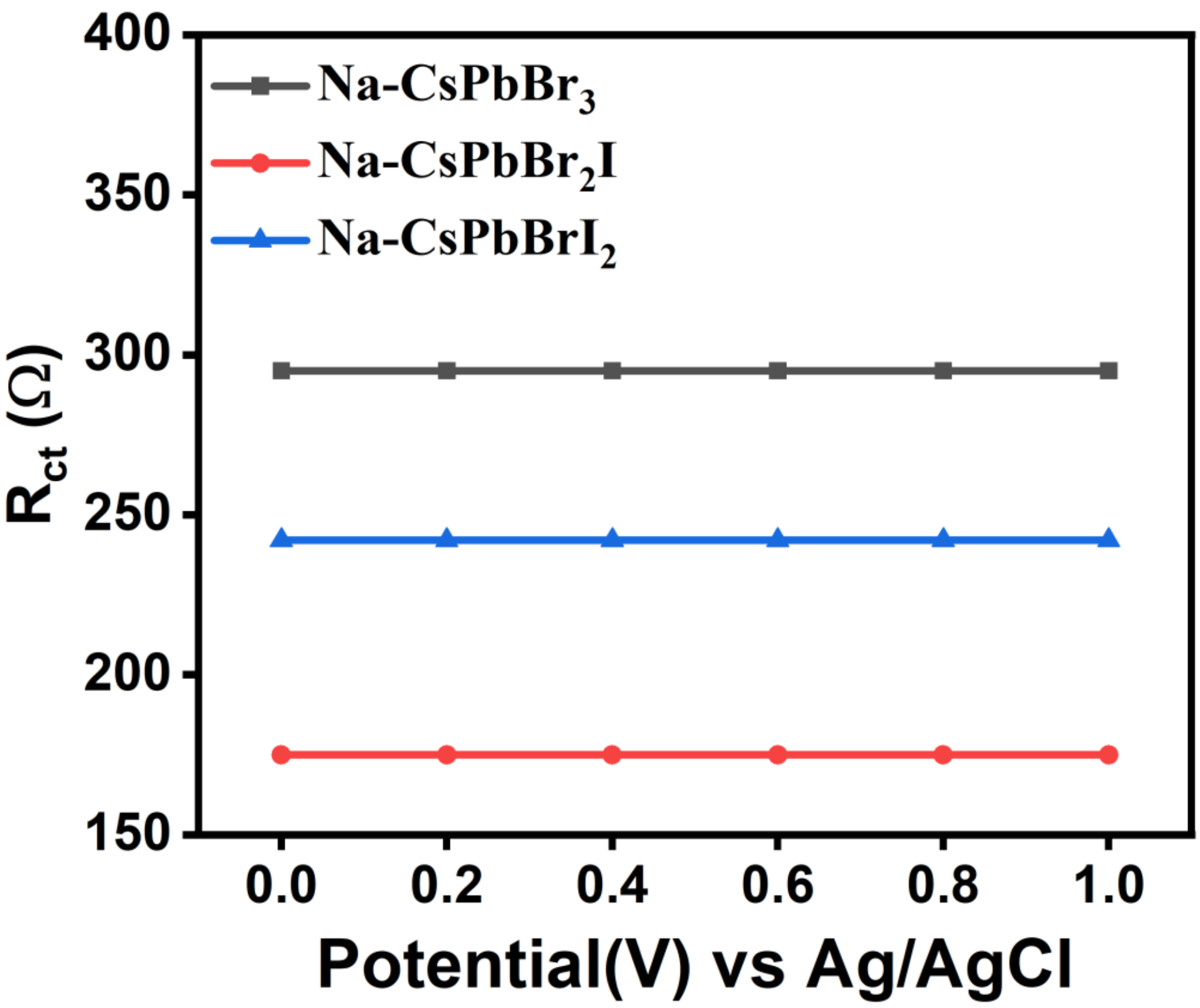

**Figure S9.** Charge-transfer resistance for each composition in voltage range from 0.0 V to 1.0 V.

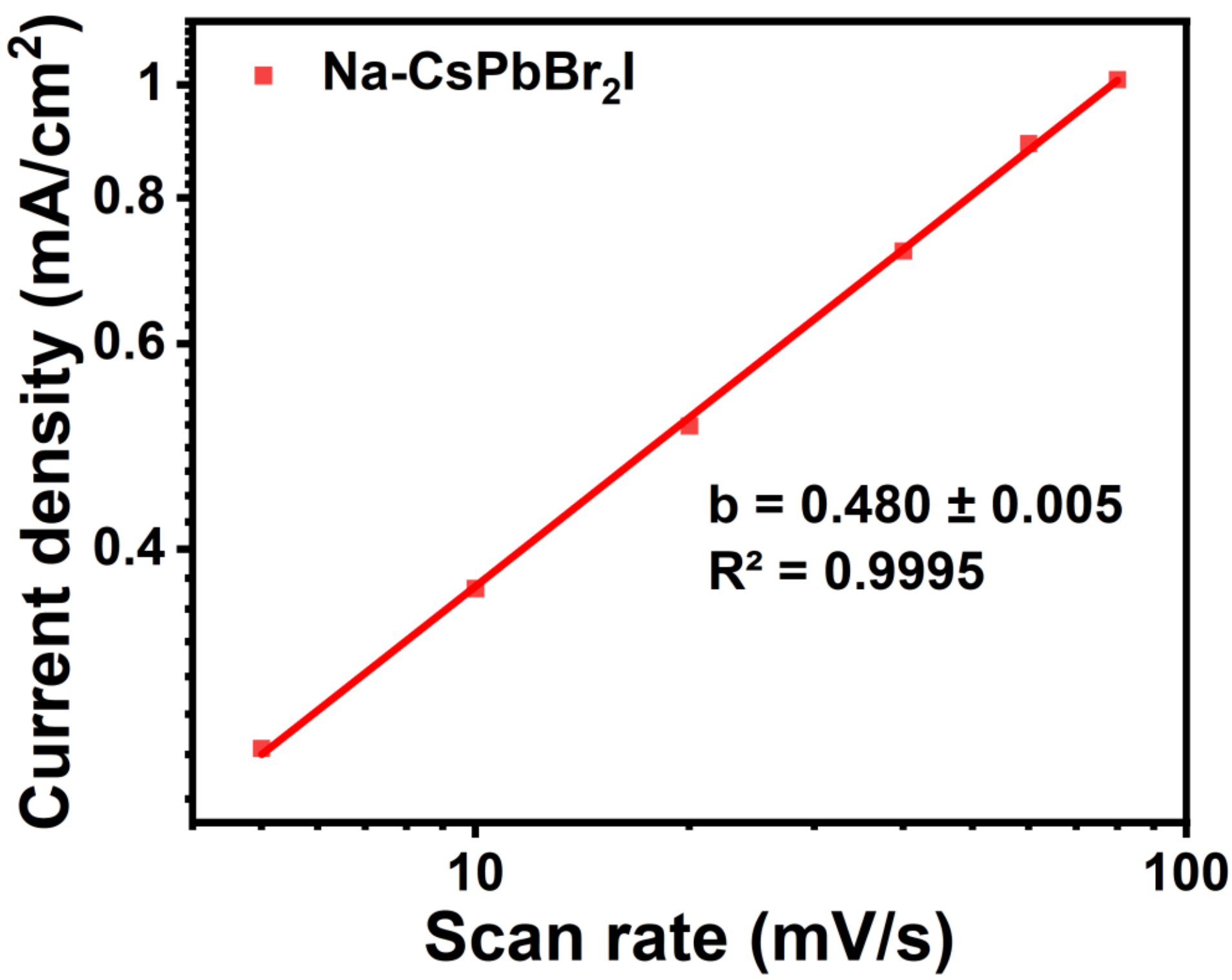

**Figure S10.** Log-log plot of current density versus scan rate at 0.2 V for Na-$CsPbBr_2I$.

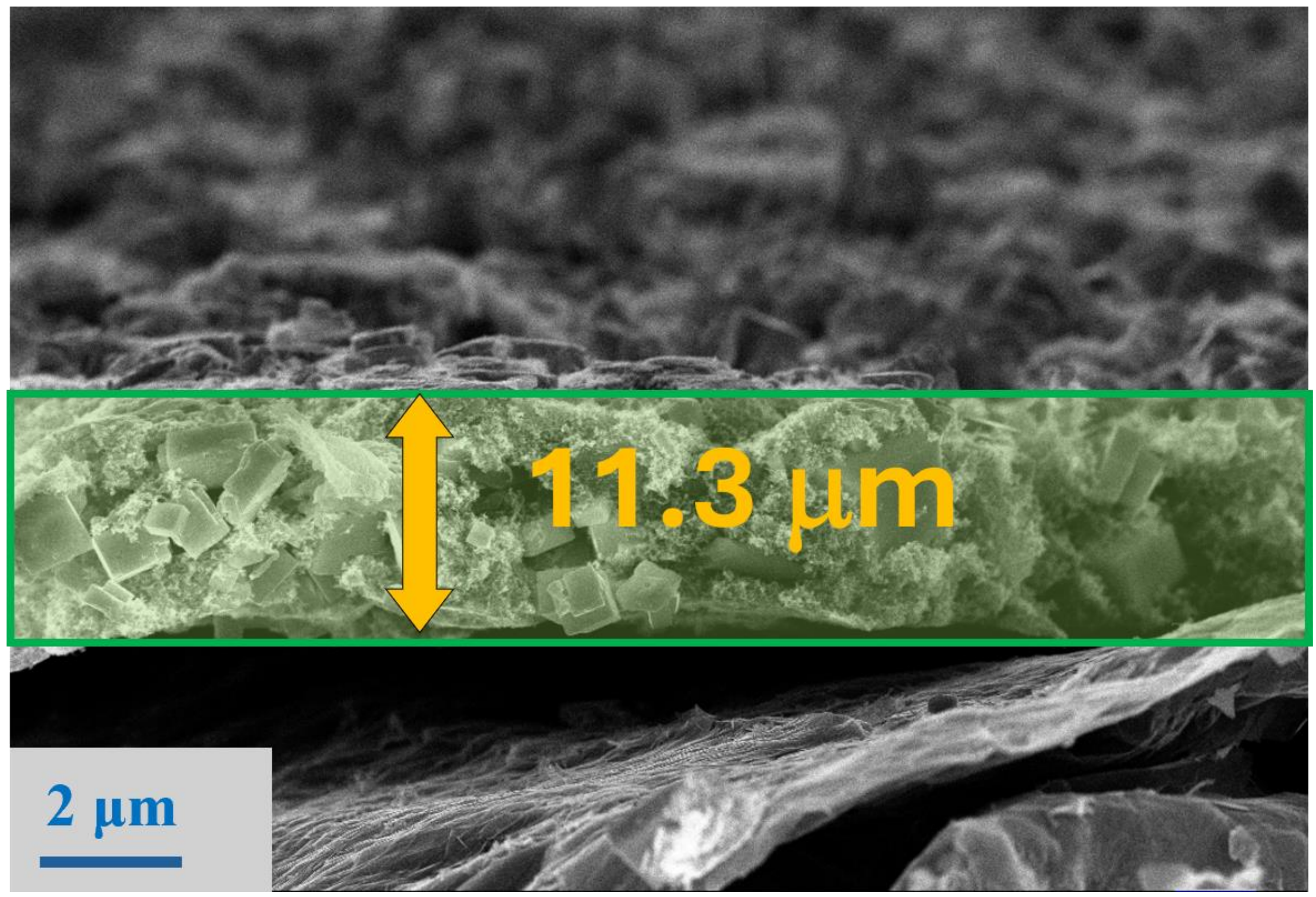

**Figure S11.** Cross-sectional FESEM image of the fabricated electrode, on the graphite substrate. The active layer average thickness is approx. 11.3 μm.